\documentclass[11pt]{article}
\usepackage{arxiv}
\usepackage[utf8]{inputenc}
\usepackage[T1]{fontenc}
\usepackage{amsmath,amssymb,amsfonts}
\usepackage{graphicx}
\usepackage{booktabs}
\usepackage{tabularx}
\usepackage{array}
\usepackage{multirow}
\usepackage{tikz}
\usetikzlibrary{arrows.meta,positioning,fit,calc}
\usepackage{placeins}
\usepackage{float}
\usepackage{capt-of}
\usepackage{longtable}
\usepackage{listings}
\lstdefinestyle{tool}{
  basicstyle=\ttfamily\scriptsize,
  breaklines=true,
  breakatwhitespace=true,
  columns=fullflexible,
  keepspaces=true,
  frame=single,
  aboveskip=6pt,
  belowskip=6pt
}
\usepackage{algorithm}
\usepackage{algpseudocode}
\usepackage{xcolor}
\usepackage{setspace}
\usepackage[numbers,sort&compress]{natbib}
\usepackage[hidelinks]{hyperref}
\usepackage{xurl}

\makeatletter
\let\flushbottom\raggedbottom
\AtBeginDocument{%
  \raggedbottom
  \def\@textbottom{\vskip \z@ plus 1fill}%
  \let\@texttop\relax
  \clubpenalty=0
  \widowpenalty=0
  \displaywidowpenalty=0
  \brokenpenalty=0
}
\makeatother

\graphicspath{{figures/}}
\newcommand{\kstar}{k^{\ast}}
\newcommand{\Estar}{E^{\ast}}

\title{A property-registry contract for retrieve-or-refuse
thermal--mechanical lattice search}
\renewcommand{\shorttitle}{Property-registry retrieve-or-refuse lattice search}

\author{
  Shaoliang Yang\thanks{Department of Mechanical Engineering, Santa Clara University, Santa Clara, CA 95053, USA.} \and
  Henry Chu\footnotemark[1] \and
  Zu Yashengjiang\footnotemark[1] \and
  Jun Wang\thanks{Corresponding author. Email: jwang22@scu.edu.}
}
\date{5 September 2026}

\begin{document}
\maketitle

\begin{abstract}
Early thermal--mechanical lattice requirements are knowledge-intensive and
often jointly unsatisfiable: an engineer asks for a cell that is light, stiff,
laterally conducting and cheap, and no cell in the library satisfies it.
A design system should say so, and say which requirement to loosen and by
how much, rather than return the nearest row.
A generative model can return a candidate even when the library holds none.
This work applies established conflict diagnosis to a catalogue of homogenised
properties.
Search over $1{,}397$ homogenised unit cells and $19$ base materials
($26{,}543$ combinations) returns a catalogue row that a second solver can
rebuild, or, for an empty feasible set of at most eight constraints,
inclusion-minimal unsatisfiable subsets (MUS) and the slack of a repair, not a
neighbour. One registry declaration generates the prompt and the evaluator, so
undeclared parsed keys cannot reach search.
On a frozen suite of 64 typed queries, retrieve-or-refuse matches min-repair
on every feasible query ($48/48$) and refuses every empty one ($16/16$) with
MUS and slack. Constraint-ignoring nearest-neighbour and penalty search
violate stated cost or density limits. On 216 empty queries, repairs printed
at three significant figures and rounded outward stay feasible in every case
($216/216$). When density and cost must be kept, a list of
minimum-cardinality repairs keeps them on $209$ of $211$ queries; a
protection-first repair keeps them on all $211$, the same as full diagnosis.
A 308-request parse benchmark is a template-text check,
not unconstrained engineer prose.
\end{abstract}


\section{Introduction}
\label{sec:intro}

An engineer specifying a heat-spreading lattice rarely begins with a target
tensor. The request is qualitative, coupled, and knowledge-bearing: light,
stiff along the length, conduct sideways, insulate through the thickness, and
do not spend silver. Over any finite library of cells and metals, the feasible
set of that request is often empty. Topology optimisation answers a different
question---the best structure for one load case---at the cost of a PDE solve
per query. Deep generative models answer a third: sample the design manifold.
Neither returns a short list of cells that can be meshed tomorrow, together
with a reason when the request cannot be met. The question this paper answers
is whether a knowledge-grounded retrieve-or-refuse loop can give an engineer
an auditable cell, or an attributable infeasibility, without generating
geometry.

That question sits in the line of knowledge-intensive engineering
support~\cite{tomiyama1994kie,gero1990prototypes,umeda1996fbs}: a formal
representation of what may be asked, a deterministic reasoner over stored
facts, and a human revision loop when the facts refuse the ask.
Language-driven inverse design has closed the interface by generation: a
sentence becomes a microstructure, a truss, or a program, and the system
returns a candidate rather than an infeasibility
explanation~\cite{kartashov2025cmame,lin2026trussgpt,makatura2025metagen}.
Graph-token reconstruction likewise generates geometry; it is not a
user-language requirement parser~\cite{khanghah2025jcise}.
Engineering-informatics neighbours are stronger on knowledge contracts and on
evaluation, and weaker on
refusal~\cite{wan2025aei,chen2026aei,liang2025aei,vyas2026aei};
Sec.~\ref{sec:related-llm} takes each in turn. Tabulated truss
catalogues~\cite{panetta2015truss} and procedural
geometry~\cite{makatura2023procmeta} retrieve stored cells rather than
emitting a sentence-conditioned sample; they do not treat an empty request as
a named conflict.

The intended claim is narrow, because several neighbouring claims are no longer
available. ``Natural language plus metamaterials'' is
taken~\cite{kartashov2025cmame,lin2026trussgpt,makatura2025metagen}. Coupled
multiphysics inverse design without language is
taken~\cite{wang2026natcomm}. Infeasibility diagnosis in natural language is
taken in operations
research~\cite{chen2024optichat,chen2025optichat,li2025moid}. Scale is not a
contribution: Lili Wang~\emph{et al.} published $404{,}355$ microstructures at
$128^3$~\cite{wang2026natcomm}; the catalogue here is $1{,}397$ cells at
$32^3$--$48^3$, by design, because every cell is a homogenisation solve and
none is a prediction.

The combination is late rather than easy, and two things kept it apart.
Generative inverse design removed the incentive to notice an empty feasible
set: a model that can always emit a candidate never has to report that none
exists, and in the largest such study the inverse-design prompts are sampled
from rows that already exist, so an unsatisfiable request never enters the
evaluation~\cite{makatura2025metagen}. Building the product is the other
cost: $1{,}397$ cells
at $32^3$--$48^3$, each a homogenisation solve and not a prediction, which
buys nothing unless refusal is the outcome the system is built to support.
Minimal conflicts and minimal repairs are settled theory in model-based
diagnosis and in cooperative database
answering~\cite{reiter1987diagnosis,junker2004quickxplain,godfrey1997mfs}.
This work applies that diagnosis to a catalogue of homogenised properties
with reproducible geometry and numerical provenance.

Two properties remain, and they are what this paper claims. The first is about
the search: when the feasible set is empty it returns why, not the nearest row.
The second is about the interface: a quantity the registry does not declare
cannot reach the evaluator as though it did. The lattice catalogue is the
instantiation that makes those two properties testable on a coupled
thermal--mechanical product. The language model is a replaceable compiler of
sentences into the registry contract; it does not retrieve, rank, or invent a
cell.

The contributions are:
\begin{enumerate}
\item A retrieve-or-refuse reasoner over the compiled query. Feasible queries
return a rank-normalised catalogue row. Infeasible queries with at most eight
remaining constraints return minimal unsatisfiable subsets, all
inclusion-minimal correction sets, a minimum-cardinality repair, and a
jointly attainable slack vector; a larger empty query is refused with that
cap named. They do not return a neighbour.
\item A property-registry contract for knowledge-grounded design search.
Entities, units, axes, provenance, admissible operators, relations and
validation rules are declared once. The prompt and the evaluator are generated
from that declaration, so the prompt and the evaluator share vocabulary,
units, and searchable keys by construction. The language model can still mis-assign role,
operator, or threshold. A parsed key the registry does not declare is
residue; a requirement the model omits is not thereby detected.
Estimated quantities are first-class.
\item An instantiation on $1{,}397$ periodic TPMS-family cells, each storing
the $3\times 3$ conductivity and $6\times 6$ stiffness from voxel
finite-element homogenisation, crossed with 19 base materials
($26{,}543$ combinations). Thermal factorisation is exact; elastic
factorisation, re-solved on every family and both modes at $\rho\approx 0.35$,
drifts by at most $1.3\%$ over $\nu\in[0.29,0.34]$ (median $0.60\%$).
\item An evaluation on six questions: parse quality as a deployment check on
template text; retrieve-or-refuse against constraint-aware min-repair on a
frozen suite of 64 typed queries; whether a printed repair stays feasible
when typed back, and whether the order of repairs changes which requirement
is kept (Sec.~\ref{sec:repair-witness}); whether the contract takes a second
solved property without a reasoner change (Sec.~\ref{sec:dstar}); four
worked briefs with MUS scored against an independent enumerator; and query
latency plus rank and refusal under mesh-scale property noise.
\end{enumerate}

\section{Related work}
\label{sec:related}

Table~\ref{tab:peers} places the present system against the design-system peers
that a reviewer will reach for. The comparison is by knowledge contract,
interaction, and what happens when the request cannot be met---not only by
whether a sentence is accepted. Close peers generate or retrieve a candidate; they do not evaluate explicit
infeasibility explanations. Related but non-design peers already refuse
(Sec.~\ref{sec:related-infeas}); what they lack is a named conflict that a
second solver can rebuild.

\subsection{Knowledge representation for design}
\label{sec:related-kr}
Portable ontologies~\cite{gruber1993ontology}, engineering
ontologies~\cite{borst1997engont}, function--behaviour schemas
\cite{gero1990prototypes,umeda1996fbs,kitamura2004functional} and
knowledge-intensive engineering~\cite{tomiyama1994kie} are the home literature
for declaring what a designer may ask. Constraint-based CAD and collaborative
design treat over-constrained specifications as objects to be diagnosed and
negotiated, not as solver failures~\cite{anderl1996constraints,lottaz2000negotiation}.
Consistency-based diagnosis of configuration knowledge bases uses the same
conflict and repair objects on product configuration rather than on a CAD
sketch~\cite{felfernig2004config}.
The registry in this paper is not a general product ontology. It is a much
smaller contract, specialised to typed search over stored effective
properties: the language interface cannot silently pass an out-of-registry
quantity to the reasoner as an evaluable key.

\subsection{Knowledge-grounded language systems in engineering informatics}
\label{sec:related-llm}
Mustapha surveys LLM use in mechanics, product design and
manufacturing~\cite{mustapha2025aei}.

Wan~\emph{et al.} compare vector-only, knowledge-graph-only and hybrid
retrieval for domain Q\&A over manufacturing knowledge graphs, with
exact-match and context precision as the metrics~\cite{wan2025aei}. That is
passage retrieval for a question that is assumed to have an answer. It is not
typed constraint search, and it cannot come back empty with a minimal
conflict.
Schema-grounded semantic parsing compiles a sentence onto a declared schema
and checks the result by execution, most visibly as text-to-{SQL}~\cite{yu2018spider}.
The parse stage here is that pattern with a generated property registry in
place of a database schema; execution is catalogue search, which may return
empty.

Chen and Bao coordinate an LLM with a genetic optimiser and finite-element
analysis for ultra-high-performance concrete beams, and report 88\% less
human effort than manual design~\cite{chen2026aei}. The evaluation is
effort-measured generation and optimisation. An unmet requirement is handled
by search, not by naming a conflict.

Liang~\emph{et al.} couple large vision--language models with topology
optimisation across 2D and 3D tasks, compare against classical BESO, and
include compression experiments~\cite{liang2025aei}. The output is a generated
layout. Physical tests qualify that layout; they do not diagnose an empty
feasible set.

Vyas~\emph{et al.} run a 46-participant study of how contextual fidelity
changes what designers think, say and do under AI-assisted engineering
design~\cite{vyas2026aei}. They measure designers. This paper does not; the
revision loop in Fig.~\ref{fig:pipeline} is an interface claim, not a user
study.

Kruiper~\emph{et al.} ground regulatory compliance in a 420-document corpus
rather than in generated prose~\cite{kruiper2024aei}. Liu~\emph{et al.}
describe a fusion AI-AD architecture from understanding to generation in
mechanical design~\cite{liu2025aei}. Duan and Wu invert lattice stiffness
targets with a GAN that also takes process parameters~\cite{duan2025aei}.
Those systems evaluate end-to-end design or retrieval quality. They generate,
retrieve passages, or optimise; they do not refuse a typed engineering query
with a minimal conflict and a slack.

\subsection{Generative inverse design}
\label{sec:related-gen}
Diffusion, variational, and transformer generators now cover trusses, porous
graphs, and voxel microstructures~\cite{kartashov2025cmame,lin2026trussgpt,khanghah2025jcise,zheng2026diffumeta}.
A transformer foundation model inverts unit-cell compressive response without
a language interface~\cite{kim2026metafo}.
Makatura~\emph{et al.} (MetaGen) give a vision--language model a
domain-specific language and a database of $153{,}263$ elastic
metamaterials~\cite{makatura2025metagen}. Inverse-design prompts are sampled
from rows that already exist, so they are feasible by construction, and the
model emits a new program. The contribution is a generated geometry, not a
refused query. Chen~\emph{et al.} (MetaSymbO) retrieve a scaffold from a
language intent and evolve it in a symbolic latent~\cite{chen2026metasymbo}.
Zhao~\emph{et al.} (AutoMS) parse a natural-language request and run
multi-agent evolutionary search over coupled mechanical and thermal
response~\cite{zhao2026automs}. Both start from language and keep searching.
Neither treats an empty feasible set as a first-class outcome.

The load-bearing property of these methods is that they can propose geometries
that are not in any catalogue. The matching cost is that the property attached
to a proposal is a network output or a surrogate, and that an infeasible
target still produces a sample. Experimental validation, where it exists, is
of fabricated generated parts~\cite{zheng2026diffumeta,lin2026trussgpt}, which
this paper does not have. Architected-material design without language already
has a long computational literature, from topology optimisation~\cite{sigmund2001smo}
of periodic media~\cite{osanov2016annrev} to catalogues that sit at
theoretical stiffness limits~\cite{berger2017nature,zheng2014science,schaedler2011science}.

\subsection{Retrieve-and-verify CAD}
\label{sec:related-retrieve}
Keyword-table retrieval from an analytic implicit library is one
retrieve-and-verify pattern. Materials-selection charts retrieve a substance
from plotted properties~\cite{gibsonashby1997}. Panetta~\emph{et al.} tabulated
$1{,}205$ cubic truss topologies for fabrication~\cite{panetta2015truss}.
Procedural graphs~\cite{makatura2023procmeta} represent and generate editable
geometry. TPMS
families themselves are a mature geometry class, with tabulated
mechanics~\cite{alketan2019aem,abueidda2019gyroid,khaderi2014ijss,maskery2017msea,kapfer2011biomaterials}
and measured or computed conductivity~\cite{catchpolesmith2019am}. Retrieval
avoids invalid geometry by construction. It cannot invent a topology that is
not stored. A chart or a family table also cannot compile a coupled qualitative
brief onto a metal--cell product, and cannot return a minimal conflict when
the brief is empty.

\subsection{Infeasibility as an object}
\label{sec:related-infeas}
Model-based diagnosis~\cite{reiter1987diagnosis,dekleer1987atms} and
preferred explanations of over-constrained problems~\cite{junker2004quickxplain}
are the classical account of minimal conflicts and minimal repairs. They
explain over-constraint. They are not attached to a language-compiled
engineering query. OptiChat names a solver-isolated irreducible infeasible
subset in natural language~\cite{chen2024optichat,chen2025optichat}. MOID
diagnoses infeasible routing models by generating trade-off solutions between
route cost and constraint violation and using language-model agents to turn
those solutions into diagnostic suggestions~\cite{li2025moid}; it
distinguishes that procedure from IIS-based repair. That
is the right prior art for attaching a language interface to a diagnosis.
Neither is a design-retrieval system, and neither searches a homogenised
catalogue whose rows can be re-solved. The language model in this paper
remains confined to parsing; the refusal notice is the deterministic MUS and
slack, not a model-written sentence.

The same objects have a database name. Cooperative answering explains a query
that returns nothing by its minimal failing subqueries, and repairs it by its
maximal succeeding subqueries~\cite{godfrey1997mfs}; those are the MUS and the
correction set under other names, and the parse stage here already borrows the
other half of that frame in schema-grounded semantic
parsing~\cite{yu2018spider}. Cooperative answering demonstrated those objects
on relational queries. This paper uses the same MUS and
correction-set objects on homogenised $\kstar$ and $\Estar$ that a second
solver can rebuild from stored geometry. Enumerating all
minimal unsatisfiable subsets is the hard part in general, and the algorithmic
literature is about doing it incrementally at scale~\cite{liffiton2008mus}; at
the eight-constraint cap used here exhaustive enumeration is exact.
At eight constraints the subset count is at most $256$; this paper reports
measured latency rather than a runtime comparison against an incremental
enumerator. A larger query is refused with the cap
named rather than approximated. Closest on the other side, Khan returns
minimal infeasible subsets as rejection certificates for satellite scheduling
and measures their soundness and stability, without a language
interface~\cite{khan2026whynot}. The certified object there is a schedule;
here it is a physical row that a second solver can rebuild and re-homogenise.

This paper takes those diagnosis objects---inclusion-minimal unsatisfiable
subsets and a minimum-cardinality repair---and attaches them to a generated
property registry and a solved metal--cell product. That combination is what
Table~\ref{tab:peers} records as refuse.

\subsection{The remaining gap}
\label{sec:related-gap}
Two properties remain after the literature above, and they are the ones
Table~\ref{tab:peers} already shows. No neighbouring \emph{design} system
generates the prompt and the evaluator from one registry declaration, so
vocabulary, units, and searchable keys cannot silently diverge. No neighbouring
design system retrieves over a solved metal--cell product and, when the
feasible set is empty, returns a named MUS and slack rather than a neighbour, a
generated candidate, or a predicted property. Those two properties are the
claims in Sec.~\ref{sec:intro}; the catalogue is the instantiation that makes
them testable, and the suite is the evidence.

\begin{table}[htbp]
\centering
\caption{Where neighbouring design systems represent engineering knowledge, generate
or retrieve, and refuse. Headline evidence is the authors' own, not a ranking.}
\label{tab:peers}
\setlength{\tabcolsep}{3.5pt}
\small
\begin{tabular}{@{}p{3.15cm}p{2.55cm}p{2.35cm}p{1.15cm}p{2.7cm}@{}}
\toprule
System & Knowledge & Mode & Refuse & Headline evidence \\
\midrule
Kartashov \& Vlassis~\cite{kartashov2025cmame} & commands & generate & no & 2D cases \\
TrussGPT~\cite{lin2026trussgpt} & mech.\ targets & generate & no & experiment \\
Khanghah~\emph{et al.}~\cite{khanghah2025jcise} & graph tokens$^{\mathrm{a}}$ & generate & no & reconstruction \\
MetaGen~\cite{makatura2025metagen} & DSL + database & generate & no & case-study re-sim. \\
MetaSymbO~\cite{chen2026metasymbo} & language + latent & retrieve+evolve & no & validity / novelty \\
AutoMS~\cite{zhao2026automs} & NL + FEA loop & evolve & no & 17 cross-physics \\
DiffuMeta~\cite{zheng2026diffumeta} & algebraic language & generate & no & fabricated tests \\
Wan~\emph{et al.}~\cite{wan2025aei} & KG + vectors & retrieve text & no & EM / precision \\
Chen \& Bao~\cite{chen2026aei} & LLM+FEA+GA & generate/opt. & no & 88\%/75\% effort \\
Liang~\emph{et al.}~\cite{liang2025aei} & LVLM priors & generate (TO) & no & 2D/3D + test \\
Vyas~\emph{et al.}~\cite{vyas2026aei} & context fidelity & mixed-init. & no & 46 participants \\
This work & generated registry & retrieve & yes & typed suite + MUS \\
\bottomrule
\end{tabular}\\[2pt]
{\raggedright\footnotesize $^{\mathrm{a}}$The LLM reconstructs graph edges; it
is not a user-facing parser. OptiChat, MOID, cooperative
answering~\cite{godfrey1997mfs} and Khan~\cite{khan2026whynot} already refuse;
they are not design-retrieval systems and are compared in
Sec.~\ref{sec:related-infeas}.\par}
\end{table}

\section{Method}
\label{sec:method}

\begin{figure}[htbp]
\centering
\includegraphics[width=0.92\textwidth]{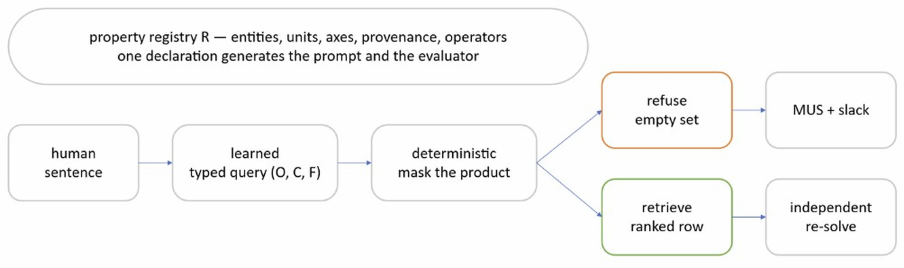}
\caption{Retrieve or refuse. The registry is the contract between language and
search. Only the parse is learned. An empty feasible set of at most eight
remaining constraints returns a minimal conflict and slack; a larger empty
query is refused with that cap named. A non-empty set returns a catalogue
row, re-homogenised on return. The engineer revises the requirements; the
system does not invent a neighbour. \ref{app:transcript} prints one
worked tool transcript.}
\label{fig:pipeline}
\end{figure}

The pipeline has four stages, drawn in Fig.~\ref{fig:pipeline}. They implement
the two properties of Sec.~\ref{sec:intro}: compiling a sentence against a
generated registry, then returning a catalogue row or a named refusal. A bad
answer is always attributable to one stage: an undeclared parsed key, a
compiled query whose feasible set is empty, a ranking among
survivors, or a stored number that fails an independent re-solve.

\subsection{Property registry as a knowledge contract}
\label{sec:registry}

This is the interface property: a quantity the registry does not declare
cannot reach the evaluator as though it did. Let $\mathcal{R}$ be a finite set of properties. Each $p\in\mathcal{R}$ is a
tuple
\begin{equation}
  p = (\mathrm{key},\;\mathrm{label},\;u,\;\kappa,\;\pi,\;\preceq,\;h),
  \label{eq:prop}
\end{equation}
where $u$ is a unit (or dimensionless), $\kappa\in\{\mathrm{geometry},
\mathrm{material},\,\mathrm{effective}\}$ is the kind, $\pi$ is provenance,
$\preceq$ is the usual engineering direction if one exists (high or low), and
$h$ is the hint string shown to the language model. Provenance is one of
\emph{solved} (voxel homogenisation of a rebuilt mask), \emph{handbook}
(nominal room-temperature wrought or standard-process constants, used for
ranking), or \emph{estimated} (a ranking surrogate that must not be presented
with the same authority as a solve).

The handbook layer is 19 metals and ceramics carrying conductivity, modulus,
density, thermal expansion, bulk price and service temperature. These are
consensus room-temperature values of the kind tabulated in the CRC Handbook of
Chemistry and Physics~\cite{haynes2014crc} for the elements and in ASM Handbook
Vol.~2~\cite{asm1990vol2} for the alloys, spot-checked against those sources and
against supplier datasheets for the named tempers; they are nominal rather than
measured here, and additively manufactured parts in particular sit below wrought
values. Gold is carried as a reference point rather than as a candidate
structural material, and is excluded from no query only because the price
constraint removes it wherever cost is stated.

The accuracy of any individual row matters less than it appears, because every
effective property factorises as a material value times a dimensionless
geometry factor. For a fixed material, a positive uniform scaling of that
material's scalar properties does not change the ranking of cells of that
metal under the paper's scalar rank-normalised rule: the cells move together.
Perturbing all 19 rows by $\pm 50\%$ leaves the ranking of all $1{,}397$
feasible cells identical (maximum rank shift $0$) under that within-metal
scalar ranking. The $94\times$ spread across the 19-row table is a
nominal-table figure; it can move a metal against another metal and against an
absolute threshold, and it is not a claim that constrained multi-objective
search cannot reorder cells. A user needing an absolute threshold should
substitute a measured value for their own feedstock, which requires no
re-solve because the geometry factors do not depend on it.

Admissible operators are $\{\le,<,\ge,>,=\}$ on numeric keys and $\{=\}$ on
categorical keys (\texttt{printable}, \texttt{symmetry}).

Properties are of three kinds, and the distinction is the knowledge content of
the catalogue:
\begin{enumerate}
\item \emph{Geometry} --- dimensionless, set by the cell: relative density
$\rho$, the conductivity ratio $k_{33}/k_{11}$, the stiffness ratio
$E_{33}/E_{11}$, symmetry class. Both ratios are positive axis ratios: a
value below one emphasises axis~1, a value above one emphasises axis~3, and
$1.0$ means equal response along those two axes, not full elastic isotropy.
\item \emph{Material} --- set by the substance: bulk price, thermal expansion,
service temperature, whether the metal is routinely printed.
\item \emph{Effective} --- the product of the two, which is the design space
the user actually wants: $k_{ii}$, $E_{ii}$, part density, specific stiffness,
cost per unit volume.
\end{enumerate}

Effective conductivity and stiffness are related to the stored geometry factors
by
\begin{align}
  k_{ii}^{\mathrm{eff}} &= k_s\,\kstar_{ii}(\mathrm{geometry}),
  \label{eq:kfactor}\\
  E_{ii}^{\mathrm{eff}} &\approx E_s\,\Estar_{ii}(\mathrm{geometry};\,\nu{=}0.3).
  \label{eq:Efactor}
\end{align}
Equation~\eqref{eq:kfactor} is an identity of the cell problem
(Sec.~\ref{sec:factor}). Equation~\eqref{eq:Efactor} is an approximation,
measured rather than assumed (Sec.~\ref{sec:factor}). Axis convention is part of the
contract, not a comment in the prompt: axes 1 and 2 are in-plane, axis 3 is
through-thickness.

The language-model prompt is generated from $\mathcal{R}$ by enumerating keys,
units, kinds and hints. The evaluator is generated from the
same set: a query that names a key not in $\mathcal{R}$ is rejected at
validation and recorded as residue, rather than ignored during scoring. A
hallucinated key is the same event. Estimated properties travel with an
explicit caveat string; permeability is registered as a Kozeny--Carman
estimate from porosity and surface area and is not used as a claimed quantity
in this paper. Adding a property is one registry entry
(Sec.~\ref{sec:dstar}). Because the prompt and
the evaluator are both generated from $\mathcal{R}$, they share vocabulary
and units by construction. That is a design property of the generator, not a
measured robustness result. It is not a claim that a parsed query
preserves role, operator, or threshold: concept F1 on the 308-request
benchmark scores property-concept overlap, not those frames.

Table~\ref{tab:registry} summarises the contract that the rest of the paper
evaluates against.

\begin{table}[htbp]
\centering
\caption{Excerpt of the property-registry contract. Provenance is solved
(homogenisation), handbook (nominal metal table), or estimated (not claimed).
The full registry has 26 queryable keys; 24 of them are searchable numeric
columns.}
\label{tab:registry}
\small
\begin{tabular}{@{}lllll@{}}
\toprule
Key & Kind & Unit & Provenance & Operators \\
\midrule
\texttt{rho} & geometry & --- & solved (isovalue bisection) & $\le,\ge$ \\
\texttt{k\_aniso} & geometry & --- & solved ($k_{33}/k_{11}$, smaller is
more directional) & $\le,\ge$ \\
\texttt{D\_11} & geometry & --- & solved ($D^{\ast}/D_0$, complete periodic
pore) & $\le,\ge$ \\
\texttt{symmetry} & geometry & --- & solved (point group of $f$) & $=$ \\
\texttt{cost\_per\_kg} & material & USD/kg & handbook & $\le,\ge$ \\
\texttt{printable} & material & --- & handbook & $=$ \\
\texttt{k\_11} & effective & W/(m\,K) & solved $\times$ handbook, exact & $\le,\ge$ \\
\texttt{E\_11} & effective & GPa & solved $\times$ handbook, $\nu{=}0.3$ & $\le,\ge$ \\
\texttt{mass\_density} & effective & kg/m$^3$ & handbook $\times$ $\rho$ & $\le,\ge$ \\
\texttt{permeability} & effective & m$^2$ & estimated, not claimed & --- \\
\bottomrule
\end{tabular}
\end{table}

\subsection{Parse}
\label{sec:parse}

The model (Gemini 3.5, JSON-schema decoding, temperature 0) maps a sentence
to a typed query
\begin{equation}
  q = (\mathcal{O},\,\mathcal{C},\,F)
  \label{eq:query}
\end{equation}
and a residue record (Table~\ref{tab:query-fields}). $\mathcal{O}$ lists
objectives (property, sense, weight); $\mathcal{C}$ lists constraint atoms
(property, operator, value); $F$ is an optional material filter (printable
only, price cap, include/exclude list). A wish (``as light as possible'') is
an objective; a numbered limit is a constraint. The model does not retrieve,
rank, or invent a cell. Residue is a parsed key not in $\mathcal{R}$, or
content the model itself reports as unmet. The validator does not compare the
sentence with the parse, so an omitted requirement is not thereby detected.
The deployed
loop prints the compiled query before a row is returned, so a mis-assigned
role or threshold is visible before the row is trusted.
\ref{app:transcript} prints one worked transcript.

\begin{table}[ht]
\centering
\caption{Parse output. Only $q$ reaches search; residue is reported, not scored.}
\label{tab:query-fields}
\small
\begin{tabular}{@{}ll@{}}
\toprule
Field & Role \\
\midrule
$\mathcal{O}$ & weighted objectives \\
$\mathcal{C}$ & constraint atoms \\
$F$ & material filter \\
\texttt{UNDERSTOOD AS} & compiled $q$, printed before retrieval \\
\texttt{CANNOT EXPRESS} & schema cannot name the ask \\
\texttt{DISCARDED} & undeclared or vacuous keys \\
\texttt{ESTIMATE ONLY} & registered estimate, not a claimed solve \\
\bottomrule
\end{tabular}
\end{table}

\subsection{Homogenisation}
\label{sec:homog}

The effective properties of a fixed volume of porous material are set by
the four factors in Table~\ref{tab:homog-factors}. Only volume fraction and
porosity layout are solved for; the solid is a handbook scalar
(Sec.~\ref{sec:factor}) and the void is empty.

\begin{table}[ht]
\centering
\caption{What enters an effective property. Only the last two rows are solved.
The void is the exact empty-pore model, not a soft inclusion.}
\label{tab:homog-factors}
\small
\begin{tabular}{@{}p{2.15cm}p{3.55cm}p{1.85cm}@{}}
\toprule
Factor & In this catalogue & Solved? \\
\midrule
Solid phase & 19 handbook metals/ceramics as $k_s$, $E_s$ & no \\
Void phase & empty pore, $k_{\mathrm{void}}=0$ & no \\
Volume fraction & $\rho\in[0.108,0.514]$ by isovalue bisection & yes \\
Porosity layout & family, mode, $f$, isovalue; voxel FE & yes \\
\bottomrule
\end{tabular}
\end{table}

The void conducts nothing. This is the exact model of an empty pore rather than
a simplification of one, and it is the limit the homogenisation literature
targets: the classical closed forms for a porous solid are stated for a
non-conducting inclusion (Maxwell~\cite{maxwell1873}; the Hashin--Shtrikman
bound at zero inclusion conductivity~\cite{hashin1962}), and cellular-solids
theory treats conduction as solid-phase dominated~\cite{gibsonashby1997}.
Methods that instead assign the void a small non-zero property do so for
numerical reasons and introduce a material that is not present; adapting
solvers to genuine zero is an active concern in the FFT literature for exactly
this reason~\cite{lucarini2021cmame}. Deleting void elements, as the voxel form
here does, is the zero-contrast treatment those methods approximate.

A separate question is what a pore that is \emph{not} empty would add. We
measure it rather than leave it open. Table~\ref{tab:airfill} reports the
shift in $\kstar$ when three cells are re-solved with still air
($0.026$\,W/mK) in the pores. The shift tracks $k_{\mathrm{air}}/k_s$ and is
appreciable only for the poor conductors. Every answer retrieve-or-refuse
returns on the worked briefs is aluminium~6061 or AlSi10Mg, so the largest
such shift reaching a result here is $+0.14\%$.

\begin{table}[ht]
\centering
\caption{Change in $\kstar$ when the empty pore is filled with still air
($0.026$\,W/mK). Three cells re-solved per metal.}
\label{tab:airfill}
\small
\begin{tabular}{@{}lc@{}}
\toprule
Solid & $\Delta\kstar$ \\
\midrule
copper & $+0.02$ to $+0.04\%$ \\
aluminium~6061 & $+0.06$ to $+0.11\%$ \\
AlSi10Mg & $+0.07$ to $+0.14\%$ \\
stainless~316L & $+0.6$ to $+1.1\%$ \\
Ti--6Al--4V & $+1.4$ to $+2.7\%$ \\
\bottomrule
\end{tabular}
\end{table}

The correction does not require rebuilding the catalogue. To first order about
an empty void
\begin{equation}
\kstar \;=\; k_s\,A \;+\; k_{\mathrm{void}}\,B ,
\label{eq:voidsplit}
\end{equation}
with $A$ the stored empty-void conductivity factor and $B$ a finite-difference
(secant) coefficient at $\varepsilon=10^{-3}$ (one extra two-phase solve per
cell with $k_s=1$ and $k_{\mathrm{void}}=\varepsilon$), not an exact
derivative at $k_{\mathrm{void}}=0$. Phase-energy field averages of a
two-phase solve at a given contrast,

\begin{align}
  A &= \rho\,\langle|E|^2\rangle_{\mathrm{s}},
  \label{eq:Afield}\\
  B &= (1-\rho)\,\langle|E|^2\rangle_{\mathrm{v}},
  \label{eq:Bfield}
\end{align}
where $E$ is the local temperature gradient normalised to the applied one,
are identities of that contrast; they are not identically the empty-void $A$
and the secant $B$. The stored pair reproduced those field averages to
$9\times 10^{-5}$ on the probe used to build $B$. The catalogue stores
$B_{11},B_{22},B_{33}$ on every searchable row ($1{,}397$ solves, $0.59$\,h).
A reader who wants a pore that is not empty therefore uses the linearisation
\eqref{eq:voidsplit} as a multiply and an add, not an exact re-solve.
The stored coefficients satisfy $B>0$ and $A+B\ge 1$ on every row and axis.
On the field-average pair the sum cannot fall below the applied field; the
stored pair inherits those inequalities numerically. $B$ is largest, up to $5.3$, exactly on
the barely-percolating cells whose largest connected solid fraction is under
$0.01$, where the gradient has nowhere to go but the pore space. Sweeping
$k_{\mathrm{void}}$ over a $200\times$ range on three cells, $B$ moves by
$0.04\%$ over the first $5\times$, $0.17\%$ over $20\times$ and $1.7\%$ over
the whole sweep: the split is a linearisation about an empty void rather than
an identity. On every cell tested $B$ exceeds the void volume fraction
$1-\rho$, by $1.35$ to $1.61\times$ --- the gyroid network at $\rho=0.30$
gives $B=1.13$ against $1-\rho=0.70$. $B$ exceeds $1-\rho$ because the gradient
is expelled from the conducting solid into the poorly conducting void. That
concentration is generic to two-phase conduction rather than a property of the
pore network: a dilute spherical pore gives $\langle|E|^2\rangle=2.25$, above
every value reached here.

The volume-fraction axis is $\rho$ from $0.108$ to $0.514$ across the
$1{,}397$ searchable rows; the isovalue bisection reaches each target to within
$0.0051$. Neither bound is a preference. At the bottom the cells stop
percolating (Table~\ref{tab:percolate}); at the top, above $\rho \approx 0.50$
a TPMS is a solid with holes rather than a lattice. Computing properties from
family, mode, frequency and isovalue is standard
practice~\cite{geers2010jcam}; the claim is the contract in front of them.

\begin{table}[ht]
\centering
\caption{Unsolvable enumerated cells ($1{,}536$ planned rows; $139$ fail).}
\label{tab:percolate}
\small
\begin{tabular}{@{}lc@{}}
\toprule
Target $\rho$ & Failure rate \\
\midrule
$0.10$--$0.15$ & $36.9\%$ \\
$0.15$--$0.20$ & $19.8\%$ \\
above $0.35$ & none \\
\bottomrule
\end{tabular}
\end{table}

Each catalogue row is a periodic voxel finite-element solve on trilinear
hexes~\cite{michel1999computational,hassani1998review,andreassen2014cms}.
The same element machinery produces $\kstar$ ($3\times 3$) and $C^{\ast}$
($6\times 6$). Void elements are removed rather than assigned a small
coefficient, which avoids a $10^{-6}$ contrast that otherwise destroys
conditioning. The resulting singular-but-consistent system is made positive
definite by pinning one node per connected solid component. Because the
fluctuation is periodic, rigid rotations are not in the null space.
Table~\ref{tab:solvers} records why the voxel form is used rather than the
usual alternatives.

\begin{table}[ht]
\centering
\caption{Solver choice for a true-zero void and a TPMS mask.}
\label{tab:solvers}
\small
\begin{tabular}{@{}>{\raggedright\arraybackslash}p{2.35cm}>{\raggedright\arraybackslash}p{5.35cm}@{}}
\toprule
Scheme & Fit to this geometry \\
\midrule
Voxel FE (used) & One routine, both physics; deleting void elements is the
zero-contrast treatment. \\
FFT~\cite{moulinec1998cmame} & Slow at infinite contrast, which is the regime
here. \\
Mori--Tanaka~\cite{mori1973acta} & Dilute ellipsoids; TPMS networks are
percolating and not dilute. \\
\bottomrule
\end{tabular}
\end{table}

Asymptotic expansion gives the same cell problem as the variational route
used here~\cite{hassani1998review}; the voxel form reuses one element routine
for both physics on a mask the geometry step already produces.

The cell is an analytic implicit surface (eight TPMS-style families, network
or sheet, integer frequency $f=(f_1,f_2,f_3)$). The frequency vector is how
many times the surface repeats along each axis inside the unit cell, so
$f=(1,1,1)$ is one period per side and $f=(1,1,3)$ is three periods through
the thickness. It is the only way anisotropy enters: equal integers give a
cubic cell whose conductivity tensor is isotropic by symmetry; unequal
integers change path length and ligament connectivity per direction. Every
directional result in this paper follows from that one choice. Integer
frequencies keep the cell exactly periodic. A second-rank tensor invariant
under the cubic group is isotropic, so no cubic cell can steer heat. The
isovalue is bisected to a target relative density, so property comparisons are
at fixed density rather than density in disguise. Working resolution follows
the shortest period,
\begin{equation}
  n =
  \begin{cases}
    32 & \text{if }\max_i f_i \le 2,\\
    48 & \text{if }\max_i f_i = 3.
  \end{cases}
  \label{eq:gridn}
\end{equation}
A row stores family, mode, $f$, isovalue, $\rho$, $\kstar$, $C^{\ast}$, and
derived directional moduli. The row \emph{is} the geometry: rebuilding the
mask from those four parameters reproduces the cell, so nothing stores a
voxel grid.

\subsection{Material--geometry factorisation}
\label{sec:factor}

For conductivity the cell problem is linear in $k_s$. Scaling the solid
conductivity divides out, the temperature field is unchanged, and
\eqref{eq:kfactor} is exact. For elasticity \eqref{eq:Efactor} is an
approximation, because $\Estar$ depends weakly on Poisson's ratio and the
catalogue computes the geometry factor once. Table~\ref{tab:nu} separates two
failures: a value error (the reported $E$) from a selection error (the wrong
cell). Sixteen cells spanning four families, both modes, and cubic and
tetragonal $f$ were re-solved at $\nu=0.15$, $0.30$ and $0.45$, which brackets
every material in the table. Rank shifts occur --- some cells stiffen with
$\nu$ while others soften --- but clustered cells reshuffle under any small
perturbation; the informative quantity is regret, how far the cell selected at
$\nu=0.30$ falls below the cell that is actually best at the true $\nu$. Over
the bulk of the table the identical cell is returned. Sec.~\ref{sec:practice}
therefore treats a returned row on the two gold briefs as a member of a
stable top-10 under $1\%$ mesh-scale noise, not as a unique optimum.
Refusal decisions are unchanged under a $5\%$ property
perturbation (Sec.~\ref{sec:stability}). Absolute stiffness is quoted with
confidence only for $\nu\in[0.29,0.34]$; the five $\nu$-extreme entries ---
silicon carbide ($0.17$), alumina ($0.22$), aluminium nitride ($0.24$), silver
($0.37$) and gold ($0.44$) --- carry that caveat on $E$ but not on cell
choice. None of them appears in the worked briefs. Conduction is unaffected:
$\nu$ does not enter the thermal cell problem.

\begin{table}[ht]
\centering
\caption{Single-$\nu$ catalogue ($\nu=0.30$) versus re-solves. Selection is
which cell is returned; value drift is the reported $E$.}
\label{tab:nu}
\small
\begin{tabular}{@{}lcc@{}}
\toprule
 & $\nu\in[0.29,0.34]$ & $[0.15,0.45]$ \\
\midrule
Selection regret & $0$ & $0.40\%$ at $\nu=0.45$ \\
Top-three set & unchanged & unchanged \\
Max rank shift & --- & $9$ of $16$ \\
$E_{11}$ value drift & $1.6\%$ & $11.8\%$ \\
$E_{33}$ value drift & $3.9\%$ & $33.5\%$ \\
$E_{33}/E_{11}$ drift & $3.5\%$ & $29.0\%$ \\
\bottomrule
\end{tabular}\\[2pt]
{\raggedright\footnotesize Separate assay at $\rho\approx 0.35$, $n=32$,
$f=(1,1,1)$: $E_{11}^{\ast}/E_s$ moves by at most $1.30\%$ (median $0.60\%$
over sixteen cells) between $\nu=0.29$ and $0.34$; $4.9\%$ at ceramic and
noble-metal $\nu$.\par}
\end{table}

The product is the reason the catalogue can be small. It is also the reason
the thermal--mechanical coupling is searchable at all. At fixed density,
geometry moves the ratio $k/E$ by a few times; changing the metal moves it by
$94\times$ on the 19-entry table (Inconel 718 at $0.055$ to silver at $5.17$, in
W/mK per GPa). Fig.~\ref{fig:ashby} shows both spreads. Inside density bands
of width $0.04$, geometry spread on aluminium 6061 is $2.1$--$3.3\times$
(bands $[0.20,0.24)$, $[0.30,0.34)$, $[0.40,0.44)$; $n=123$, $144$, $156$).
Those band ratios are computed from the same catalogue the search uses. The
metal is the stronger lever on exactly the coupling that makes the problem
hard, which is why the search is over pairs rather than over cell geometries.

\subsection{Search, ranking, and refusal}
\label{sec:search}

This is the search property: when the feasible set is empty the search
returns why, not the nearest row. Candidates are the Cartesian product of usable cells and metals that pass $F$.
Algorithm~\ref{alg:search} compiles $\mathcal{C}$ into Boolean masks on that
product. Vacuous constraints (those already satisfied by every finite value)
are dropped and reported, so a $\mathrm{CTE}\ge 0$ request is not silently
discarded. If the surviving set is non-empty, candidates are ranked by a
weighted sum of rank-normalised objectives. Rank normalisation is used because
min--max scaling on a heavy-tailed property (conductivity spans $0.19$ to
$201$\,W/mK) collapses the objective: half the catalogue would score above
$0.9$ on ``minimise $k$,'' and the paired objective would decide alone. The
number of non-dominated survivors is reported as a Pareto count of available
trade-off alternatives.

If the surviving set is empty, the search does not return the nearest rows.
Algorithm~\ref{alg:mus} enumerates inclusion-minimal unsatisfiable subsets
(MUS) of $\mathcal{C}$~\cite{reiter1987diagnosis,junker2004quickxplain}. A
minimal correction set (MCS) is a hitting set of the MUS family: dropping
those constraints restores a non-empty set. All inclusion-minimal MCS are
reported; the minimum-cardinality members are the cheapest repairs in the
number of constraints. For a singleton MCS the slack is the best value of
that property among rows that satisfy the others. For an MCS of size greater
than one, independent per-constraint extrema need not be jointly attainable,
so the reasoner returns one row of the restored set that minimises the sum of
scale-normalised violations of the dropped constraints. Each constraint is an
atom $(\mathrm{property}, \mathrm{operator}, \mathrm{value})$, not a property
name: two bounds on one key are distinct members of $\mathcal{C}$. MUS
enumeration is exponential in $|\mathcal{C}|$ and is capped at eight
constraints~\cite{liffiton2008mus}; a larger empty query is refused with that
cap named rather than diagnosed. Below the cap, exhaustive enumeration is
exact. Leave-one-out
binding---the constraints whose individual removal restores candidates---is
the special case in which every MUS is hit by a singleton. The deployed
search enumerates MUS and MCS; Sec.~\ref{sec:refusal} measures where
leave-one-out disagrees.

This diagnosis is a property of the \emph{query and the catalogue}, not of the
language model.

\begin{algorithm}[t]
\small
\caption{Compile and search}
\label{alg:search}
\begin{algorithmic}[1]
\Require catalogue product $X$, query $q=(\mathcal{O},\mathcal{C},F)$, registry $\mathcal{R}$
\Ensure ranked rows, or empty with diagnosis
\State drop any $c\in\mathcal{C}$ that holds for every finite value in $X$; record it
\State reject keys not in $\mathcal{R}$; record them as residue
\State $M_F \gets$ rows of $X$ that pass material filter $F$
\ForAll{$c\in\mathcal{C}$}
  \State $M_c \gets \{x\in X: x\text{ satisfies }c\}$
\EndFor
\State $K \gets M_F \cap \bigcap_c M_c$
\If{$K=\emptyset$}
  \If{$|\mathcal{C}|>8$}
    \State \Return empty with the diagnosis cap named; do not enumerate
  \EndIf
  \State \Return $\mathrm{Diagnose}(\mathcal{C},\{M_c\},M_F)$
\EndIf
\State rank $K$ by weighted rank-normalised $\mathcal{O}$; report $|K|$ and Pareto count
\State \Return top-$k$ rows
\end{algorithmic}
\end{algorithm}

\begin{algorithm}[t]
\small
\caption{Diagnose an empty feasible set}
\label{alg:mus}
\begin{algorithmic}[1]
\Require constraint atoms $\mathcal{C}$ (each $c$ is property, operator, value), masks $\{M_c\}$, material mask $M_F$
\Ensure MUS family, MCS, minimum-cardinality MCS, repair vectors
\If{$|\mathcal{C}|>8$}
  \State \Return empty; name the cap
\EndIf
\State $\mathcal{U} \gets \{S\subseteq\mathcal{C}: M_F\cap\bigcap_{c\in S}M_c=\emptyset\}$
\State $\mathrm{MUS} \gets \{S\in\mathcal{U}: \text{no }T\in\mathcal{U}\text{ has }T\subsetneq S\}$
\State $\mathrm{MCS} \gets$ inclusion-minimal hitting sets of $\mathrm{MUS}$
\State $\mathrm{MCS}^{\ast} \gets \{S\in\mathrm{MCS}: |S|\text{ is minimal}\}$
\ForAll{$S\in\mathrm{MCS}$}
  \State $X' \gets M_F\cap\bigcap_{c\notin S}M_c$
  \If{$|S|=1$}
    \State slack$(S) \gets$ best value of the unique $c\in S$ on $X'$
  \Else
    \State slack$(S) \gets$ row of $X'$ minimising scale-normalised violation of $S$
  \EndIf
\EndFor
\State \Return MUS, MCS, $\mathrm{MCS}^{\ast}$, slacks
\end{algorithmic}
\end{algorithm}

\subsection{Independent re-homogenisation}
\label{sec:verify}

A returned row is checked in three steps, not as a physical experiment:
\begin{enumerate}
\item Rebuild the mask from family, mode, frequency, and isovalue.
\item Re-solve with matrix-free conjugate gradient on GPU. The GPU path
projects out the free constant and shares no assembled-sparse code with the
CPU catalogue builder.
\item Compare the rebuilt $\rho$ to the stored value. Isovalues were restored
to full precision after a rounding that reached $3.9\%$ density error; every
feasible row rebuilds its stored $\rho$ within $10^{-6}$.
\end{enumerate}
Agreement is evidence about the discrete problem, not about the physical
cell. It does not replace mesh-convergence checks
(Sec.~\ref{sec:catalogue}) and it is not an experiment. The interactive tool
re-solves the top row by default; that step can be switched off. The table in
Sec.~\ref{sec:checks} is a sampled catalogue check of that second
implementation, not a claim that every ranked row in this paper was re-solved
at write time.

\section{Catalogue and numerical checks}
\label{sec:catalogue}

\subsection{Scope}
\label{sec:scope}

\begin{figure}[htbp]
\centering
\includegraphics[width=\columnwidth]{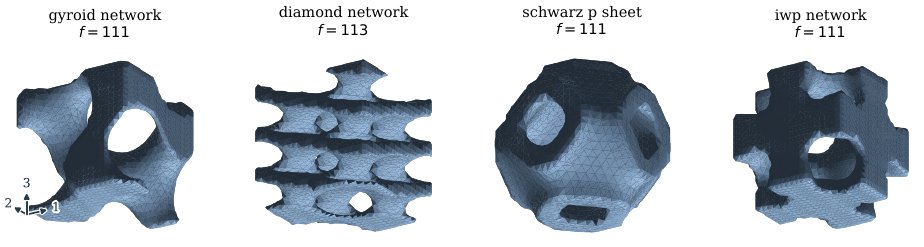}
\caption{Four catalogue cells, rebuilt from the stored parameters. Left to
right: gyroid network $f{=}111$; diamond network $f{=}113$ (tetragonal, the
steering mechanism); Schwarz P sheet $f{=}111$; IWP network $f{=}111$. The
triad on the first panel fixes the coordinate convention used throughout:
axes $1$ and $2$ are in-plane, axis $3$ is through-thickness, and all four
cells are drawn in the same orientation. Every directional quantity in this
paper ($k_{33}/k_{11}$, $E_{33}/E_{11}$, the steering result) is referred to
these axes.}
\label{fig:cells}
\end{figure}

The planned grid is Table~\ref{tab:grid}. After discarding non-percolating or
failed solves, $1{,}397$ rows ($90.9\%$) are usable. Of those, $31$ have
largest-connected-solid fraction below $0.01$ (minimum $7.5\times 10^{-4}$),
all sheet-mode at low density: more than $99\%$ of the solid is floating
islands that carry no load. The discard test keys on percolation of the
effective property, not on connectivity; \texttt{conn\_frac} is stored on
every row but is not a queryable registry key. An insulation query that
minimises $k_{11}$ can return those rows: four of the five lowest-$k_{11}$
product rows have \texttt{conn\_frac} below $0.03$ (the absolute lowest does
not). Families are gyroid, Schwarz P, diamond, IWP, Neovius, Fischer--Koch S,
FRD, and split-P. Frequency vectors cover cubic $(1,1,1)$, $(2,2,2)$;
tetragonal $(1,1,2)$, $(1,2,2)$, $(1,1,3)$, $(1,3,3)$, $(2,2,3)$; and
orthorhombic $(1,2,3)$. Fig.~\ref{fig:cells} shows four representative cells.
Material values are nominal room-temperature wrought or standard-process
figures, used for ranking. They are not a substitute for a handbook when a
number enters a drawing, and additively manufactured parts typically fall
below them.

\begin{table}[ht]
\centering
\caption{Enumerated catalogue before and after percolation discard.}
\label{tab:grid}
\small
\begin{tabular}{@{}lc@{}}
\toprule
Axis & Count \\
\midrule
Families $\times$ modes $\times$ $f$ $\times$ target $\rho$ &
$8\times 2\times 8\times 12$ \\
Target $\rho$ interval & $[0.12,0.50]$ \\
Enumerated cells & $1{,}536$ \\
Usable rows & $1{,}397$ ($90.9\%$) \\
Metals / product & $19$ / $26{,}543$ \\
\bottomrule
\end{tabular}
\end{table}

\paragraph{Can these cells be made?} The catalogue is dimensionless, so a
cell has no wall thickness until a size is chosen for it, and printability
is therefore a constraint on that choice rather than a property of the
library. To make the choice checkable rather than assumed, the registry
carries \texttt{min\_feature}: the mean wall or ligament thickness at a
stated cell size, estimated from relative density and specific surface as
\begin{equation}
  t = C\rho/S_v,\qquad C=2\text{ (sheet)},\quad C=4\text{ (network)}.
  \label{eq:minfeat}
\end{equation}
It is an estimate, not a measured minimum, and is flagged as such in the
registry alongside permeability. Table~\ref{tab:print} records the implied
thicknesses at two cell sizes.

\begin{table}[ht]
\centering
\caption{Estimated \texttt{min\_feature} on the $1{,}397$ feasible rows.
Micro LPBF has printed TPMS walls of $100\,\mu$m~\cite{qu2021micro}; the
$0.2$\,mm floor is a chosen screening threshold at a $10$\,mm cell, not a
universal machine limit.}
\label{tab:print}
\small
\begin{tabular}{@{}lccc@{}}
\toprule
Cell size & Median $t$ & Fraction $<0.4$\,mm & Fraction $<0.2$\,mm \\
\midrule
$10$\,mm & $0.91$\,mm & $22.5\%$ & $2.7\%$ \\
$20$\,mm & --- & --- & none below $0.3$\,mm \\
\bottomrule
\end{tabular}
\end{table}

The thinnest rows are all high-frequency sheet cells at low density. A user
who needs a printable part constrains \texttt{min\_feature} at the size they
intend to build; the search enforces it as any other constraint, including
refusing when it cannot be met together with the rest of the request.

This is also the boundary of what the catalogue claims. The stored geometry
factors $\kstar$ and $\Estar$, and the dimensionless axis ratios, are
unchanged under uniform scaling of the cell at fixed material constants;
effective $k_{ii}$ and $E_{ii}$ inherit that scaling through the factorisation,
while permeability, if used, would scale with the square of the geometric
length. \texttt{min\_feature} is a length and scales with the cell. Printability is therefore a verdict
about a chosen size rather than a property of a cell: the $2.7\%$ of rows that
fall below $0.2$\,mm at a $10$\,mm cell all clear that floor at a $20$\,mm one.
Nothing here is evidence about a particular application scale.

\subsection{Correctness of the discrete problem}
\label{sec:checks}

The CPU solver is checked on cases with known answers, all on the same element
and pinning machinery used for the catalogue:

\begin{enumerate}
\item Fully solid cell: $\kstar = k_s I$ to $10^{-15}$; $C^{\ast}=C_{\mathrm{base}}$
to $10^{-14}$.
\item Layered solid/void slabs, volume fraction $\phi=1/2$:
$\kstar=\mathrm{diag}(0,\phi,\phi)$ exactly, which is the test the fully-solid
case cannot reach; $C_{11}=0$ across the void.
\item Isolated non-percolating island: $\kstar=0$.
\item Hashin--Shtrikman upper bound for a solid/void composite,
$\kstar/k_s \le 2\rho/(3-\rho)$~\cite{hashin1962}, on gyroid, Schwarz P,
diamond, and IWP at three densities: never violated, worst ratio $0.83$. The
same algebraic expression is Maxwell's dilute result for insulating
spheres~\cite{maxwell1873}. The two statements are one bound, not two
independent validations. Sheet-type TPMS cells sit close to it, which matches
tabulated TPMS conductivity~\cite{catchpolesmith2019am}.
\item Cubic gyroid: conductivity isotropic to $3\times 10^{-15}$.
\item Two-material layered slabs, the non-degenerate resistance-network
case. Every check above that involves a void is degenerate on the series
side, because with $k_{\mathrm{void}}=0$ the harmonic mean collapses to
zero and exercises the null-space handling rather than the physics. A
second solver path that assembles over all elements with a per-element
conductivity (used only for validation, not to build the catalogue)
reproduces both closed forms for copper/stainless layers at four volume
fractions: across the layers the harmonic mean and along them the
arithmetic mean, each to a worst relative error of $8\times 10^{-14}$.
\item Maxwell dilute-sphere limit. A single spherical void at fractions
$f=0.012$ to $0.091$ agrees with $\kstar/k_s = 2(1-f)/(2+f)$ to within
$0.10\%$. At these fractions the residual is dominated by the staircased
voxel sphere rather than by the first-order truncation, and it falls by
$2.0\times$ when the grid is refined from $n=32$ to $n=80$ at fixed
geometry, which is the signature that it is discretisation and not a
solver error.
\item Wiener bracketing on fifteen real cells across five families and
three densities: $0 \le \kstar \le \rho\,k_s$ on every axis, tightest
ratio $0.68$.
\item Two-material laminate against the exact Backus solution, the elastic
counterpart of the resistance-network test. A void makes every elastic case
degenerate on the compliant side, so the same second-phase treatment is
applied to the $24\times 24$ element matrix; it is linear in the Lam\'e
constants, so $k_e = \lambda A + \mu B$ with $A$ and $B$ geometry-only and a
per-element pair of coefficients separates the phases. Steel and aluminium
layers at four volume fractions and all three layering axes reproduce the
exact $6\times 6$ to a worst relative error of $4.0\times 10^{-15}$. Unlike a
bound, this fixes all twenty-one independent components. The closed form is
itself verified first: a one-material laminate returns that material to
$3\times 10^{-16}$, and the transverse-isotropy identity
$C_{22}-C_{23}=2C_{44}$ holds to the same order.
\end{enumerate}

Sixteen such checks run as one suite and all pass; the suite ships with
the code so a reader can re-run it. Several of these have closed forms and
need no second solver, which makes the ladder cheaper to reproduce than a
commercial-code comparison and independent of anyone else's licence. A
different mesh of a catalogue cell is Sec.~\ref{sec:openfem}.

A sampled cross-implementation check (matrix-free GPU versus assembled CPU,
identical masks) is the verification of the stored numbers against a second
code path. Table~\ref{tab:verify} reports a 60-row thermal-plus-elastic sample
(seed 0) from the feasible catalogue. All sixty rows used the independent GPU
backend. Worst relative discrepancy $3.02\times 10^{-14}$; median
$4.2\times 10^{-15}$. Every row is inside the $10^{-6}$ agreement tolerance by
several orders of magnitude.

\begin{table}[H]
\centering
\caption{Sampled GPU versus CPU re-homogenisation of catalogue rows (seed 0).
Agreement is on the discrete problem, not the physical cell.}
\label{tab:verify}
\small
\begin{tabular}{@{}lcccc@{}}
\toprule
$n$ & backend & independent & worst rel.\ err. & all within $10^{-6}$ \\
\midrule
60 & GPU, matrix-free & yes & $3.02\times 10^{-14}$ & $60/60$ \\
\bottomrule
\end{tabular}
\end{table}

\subsection{Discretisation residual of the working grid}
\label{sec:mesh}

Mesh convergence is not uniform. The probe is a full factorial at $n=32,48,64$:
five families $\times$ two modes $\times$ $\rho\in\{0.25,0.35\}$ $\times$ three
frequency vectors, sixty cells, ten per density--frequency pair.
Table~\ref{tab:mesh} reports residuals of the working grid against $n=64$.
Relative density separates the residual; the frequency vector does not. Those
cells remain in the catalogue; the limitation belongs next to any quoted
median. Sec.~\ref{sec:stability} asks the operational question that a median
residual does not: whether a $1\%$ property perturbation, the scale of the
median mesh residual, changes the refuse/answer decision or the top-ranked
row.

\begin{table}[ht]
\centering
\caption{Working-grid residual against $n=64$ (sixty-cell factorial).}
\label{tab:mesh}
\small
\begin{tabular}{@{}lc@{}}
\toprule
Slice & Residual \\
\midrule
Median $\kstar$ / $\Estar$ / $k_{33}/k_{11}$ & $1.6\%$ / $2.4\%$ / $0.56\%$ \\
Median $\kstar$ at $\rho=0.25$ / $0.35$ & $2.5\%$ / $1.3\%$ \\
Worst $\kstar$ at $\rho=0.25$ / $0.35$ & $21.1\%$ / $2.9\%$ \\
Median $\kstar$ at $f=(1,1,1)$, $(1,1,3)$, $(1,2,3)$ & $0.9\%$, $2.1\%$, $2.0\%$ \\
Cells with $\kstar>3\%$ & $13/60$, all at $\rho=0.25$ \\
\bottomrule
\end{tabular}
\end{table}

\subsection{Second-library finite-element check}
\label{sec:openfem}

The closed-form ladder and the GPU path share the voxel occupancy and the
catalogue assembler. They do not test a different mesh of a real cell.
Further checks use \texttt{scikit-fem}~\cite{gustafsson2020skfem}, which
shares no assembled-sparse code with the CPU solver: two single-cube runs
on a tetrahedral mesh of the exported surface, six periodic elastic cells,
and four periodic conduction cells.

Watertight surface meshes of the Schwarz~P sheet at $\rho=0.30$ and the
IWP network at $\rho=0.40$ are filled with linear tetrahedra~\cite{si2015tetgen}
and solved as a single cube: hot and cold faces with insulated sides for
conductivity, uniaxial load with free sides for Young's modulus. Apparent
conductivity differs from the voxel cube by $+1.43\%$ and $+0.67\%$. The
elastic cube differs by $+6.8\%$ and $+2.1\%$; energy and face reaction
agree. That cube is not catalogue $E_{11}$. A free sheet is not held by
periodic neighbours, so both voxel and tetrahedral cubes sit below $E_{11}$
(Schwarz~P by $25\%$, IWP by $2.5\%$).

Catalogue $E_{11}$ is the periodic problem: the stored occupancy, a
periodic fluctuation, $C^{\ast}$ from the energy, and $E_{11}=1/S_{11}$.
Trilinear hexes in \texttt{scikit-fem} on that occupancy recover the
stored figure (Table~\ref{tab:openfem}). That is a second library on the
same discrete problem. The gyroid $0.03\%$ matches a re-solve of the
rebuilt mask. The same pipeline returns $E=1$ on a solid cube and matches
the Backus steel/aluminium $6\times 6$ to $1.2\times 10^{-15}$. Occupancy
and driver remain those of the catalogue. The check does not need a
commercial licence.

\begin{table}[htbp]
\centering
\caption{Catalogue $E_{11}$ (periodic, $E=1$, $\nu=0.30$). Voxel is the
stored number. Hex is \texttt{scikit-fem} on the same occupancy and the
same element family. Absolute relative difference is
$\lvert$hex $-$ voxel$\rvert/\lvert$voxel$\rvert$. STL
cube checks are in the text, not this table.}
\label{tab:openfem}
\small
\begin{tabular}{@{}lcccc@{}}
\toprule
cell & $k_{33}/k_{11}$ & voxel $E_{11}$ & scikit-fem hex & hex vs voxel \\
\midrule
gyroid network, $\rho=0.30$ & $1.000$ & $0.07004$ & $0.07006$ & $0.03\%$ \\
Schwarz P sheet, $\rho=0.30$ & $1.000$ & $0.08786$ & $0.08786$ & $2\times 10^{-6}$ \\
IWP network, $\rho=0.40$ & $1.000$ & $0.21106$ & $0.21106$ & $7\times 10^{-7}$ \\
gyroid network, $f{=}(1,1,2)$ & $0.489$ & $0.08897$ & $0.08897$ & $1.3\times 10^{-6}$ \\
diamond network, $f{=}(1,1,3)$ & $0.178$ & $0.16464$ & $0.16464$ & $8.3\times 10^{-7}$ \\
diamond network, $f{=}(1,2,3)$ & $0.312$ & $0.24204$ & $0.24204$ & $1.4\times 10^{-6}$ \\
\bottomrule
\end{tabular}
\end{table}

\subsection{Second-library check of the periodic conduction problem}
\label{sec:openfem-thermal}

The comparison above is elastic. Conduction is checked the same way, and it is
the physics the steering result rests on. The same occupancy is handed to
\texttt{scikit-fem}, which assembles its own Laplacian, imposes periodicity
through its own node-merging projection, and solves directly rather than by
preconditioned conjugate gradient. Agreement is closer than in the elastic
case, at a worst component difference of $1.4\times 10^{-13}$ over four cells,
because the conduction problem is scalar and the two codes reduce to the same
discrete system.

\begin{table}[htbp]
\centering
\caption{Periodic $\kstar$, \texttt{scikit-fem} against the catalogue solver on
the same occupancy, $k_s=1$. The last row is the cell behind the steering
result: an independent library returns the same $k_{33}/k_{11}$.}
\label{tab:openfem-thermal}
\small
\begin{tabular}{@{}lccc@{}}
\toprule
cell & our $k_{33}/k_{11}$ & scikit-fem & worst component \\
\midrule
gyroid network, $f{=}(1,1,1)$ & $1.0000$ & $1.0000$ & $1.5\times 10^{-14}$ \\
Schwarz P sheet, $f{=}(1,1,1)$ & $1.0002$ & $1.0002$ & $1.2\times 10^{-14}$ \\
IWP network, $f{=}(1,1,1)$ & $1.0000$ & $1.0000$ & $1.6\times 10^{-14}$ \\
diamond network, $f{=}(1,1,3)$ & $0.1780$ & $0.1780$ & $1.4\times 10^{-13}$ \\
\bottomrule
\end{tabular}
\end{table}

The three anisotropic rows in Table~\ref{tab:openfem} and the last row of
Table~\ref{tab:openfem-thermal} matter because the geometry carrying the
directional claim is checked in an independent code, in both physics, and the
ratio is reproduced rather than merely the magnitudes.

\subsection{What the catalogue can reach}
\label{sec:reach}

\begin{figure}[t]
\centering
\includegraphics[width=0.76\textwidth,height=0.20\textheight,keepaspectratio]{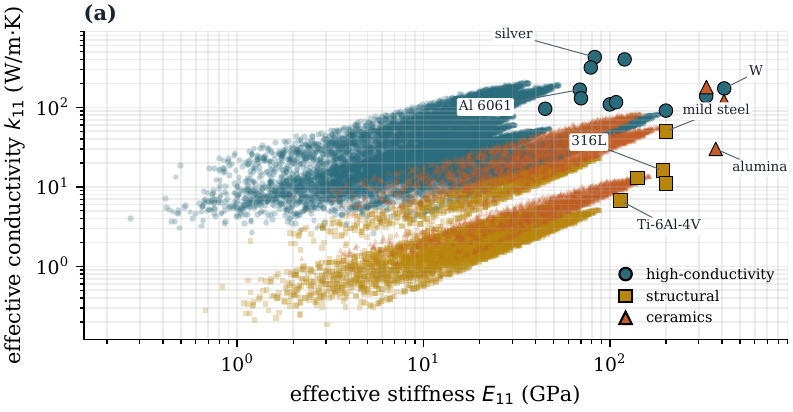}\\[2pt]
\includegraphics[width=0.76\textwidth,height=0.13\textheight,keepaspectratio]{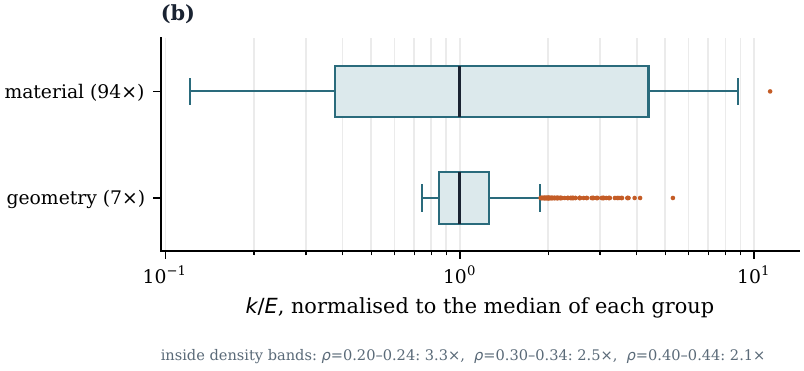}
\caption{(a)~Effective conductivity against effective stiffness for all
$26{,}543$ combinations. Clouds use colour and marker shape by material class
(circles: high-conductivity metals; squares: structural metals; triangles:
ceramics); bulk metals are larger markers with a dark edge. Labels are a subset, to avoid overlap.
(b)~Spread of the ratio $k/E$. Material choice over 19 dense solids:
$94\times$. Geometry at one metal (aluminium 6061), over the whole density
range: $7\times$ globally, $2.1$--$3.3\times$ inside density bands of width
$0.04$. The two levers are not interchangeable.}
\label{fig:ashby}
\vspace{6pt}
\includegraphics[width=0.80\textwidth,height=0.20\textheight,keepaspectratio]{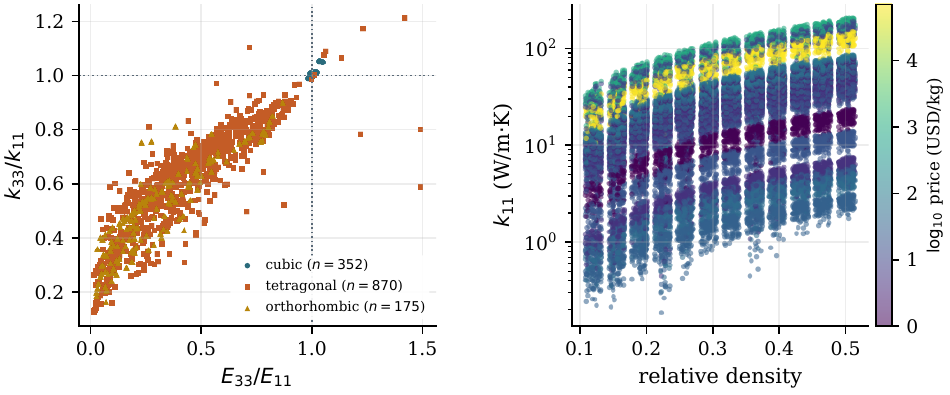}
\caption{Axes as in Fig.~\ref{fig:cells}: $1,2$ in-plane, $3$
through-thickness. Left: conductivity ratio versus stiffness ratio, coloured
by symmetry class.
Cubic cells cluster at $(1,1)$ in $k_{33}/k_{11}$ up to discretisation
(worst $k_{33}/k_{11}=0.897$ at $n=32$). Right: the same
conductivity is available at many densities and prices because the search
includes the metal.}
\label{fig:attainable}
\end{figure}

Fig.~\ref{fig:ashby} is the design space on axes an engineer already
uses~\cite{gibsonashby1997}. Each metal is a point at the top right of its own
cloud; porosity drags the point down a path. Choosing a different metal is a
near-perpendicular move. Combinations that neither a dense metal nor a
single-material lattice can reach sit in the product.

The left panel of Fig.~\ref{fig:attainable} is the symmetry obstruction in
numbers. Cubic cells sit at $(k_{33}/k_{11},\,E_{33}/E_{11})=(1,1)$ up to
discretisation: of $352$ cubic geometries, five exceed $1\%$ deviation in
$k_{33}/k_{11}$ (worst $0.897$, a $10.3\%$ miss) and seven in
$E_{33}/E_{11}$ (worst $0.886$), all at $n=32$. Directional heat exists only off that point, and only because the frequency
vector was allowed to break the cube. The right panel shows that a given
conductivity is reachable at many densities and many prices: the material
table, not a finer geometry sweep, is what opens that axis.

\section{Evaluation}
\label{sec:eval}

Six questions, in the order the argument needs them: is the language compiler
still the bottleneck on the vocabulary the registry defines; does
retrieve-or-refuse differ from constraint-aware min-repair, and from
constraint-ignoring always-answer search, on a frozen typed suite; does a
printed repair stay feasible when typed back, and does the order of repairs
change which requirement is kept (Sec.~\ref{sec:repair-witness}); does the
contract take a second solved property without a reasoner change
(Sec.~\ref{sec:dstar}); does the deployed diagnosis name the right
conflict when the feasible set is empty, including three-constraint cases;
and does mesh-scale noise flip refusal or rank. Stored-number correctness was
Sec.~\ref{sec:checks}. Sec.~\ref{sec:baselines} defines the policies and
reports the suite; Sec.~\ref{sec:repair-witness} the printed repairs and
revision task; Sec.~\ref{sec:dstar} the second property; Sec.~\ref{sec:briefs}
the four briefs. All search timings below are on the $26{,}543$-row
product; median compile-and-search latency is $2.7$\,ms on queries with at
most three constraints. A scaling probe of 16 draws at each of
$|\mathcal{C}|=1$--$6$ stayed under $7$\,ms at six constraints, where every
draw was empty and diagnosed. Diagnosis time grows with $|\mathcal{C}|$;
the deployed enumerator is capped at eight constraints. The opening heat-spreader
request compiles to five constraints, under that cap. Extending the probe to
$|\mathcal{C}|=7$--$8$ (16 draws) and $9$--$12$ (8 draws) kept median
empty-query time under $5$\,ms; queries with more than eight remaining
non-vacuous constraints are refused with the cap named.

Independently authored engineer prose was not collected. The typed briefs in
Sec.~\ref{sec:briefs} are physically motivated and were not constructed from
leave-one-out labels; they are not a substitute for a human-use study
\cite{vyas2026aei}.

\subsection{The language stage is a deployment check}
\label{sec:parseeval}

Table~\ref{tab:parse} is a deployment check, not a modelling result; the
claims of this paper are Secs.~\ref{sec:baselines}--\ref{sec:stability}. Gemini
3.5 Flash and Flash-Lite are the two model SKUs used. A 308-request benchmark
(seven categories, 44 items each, paired literal and paraphrased, all texts
unique) scores concept F1 of the parsed property set against a gold set.
That score is vocabulary overlap. It does not measure whether a property was
an objective or a constraint, nor operator, threshold, or unit. Key-level
agreement is reported alongside it; exact semantic-frame scoring is defined
in the evaluator for gold items that list constraint atoms, and was not the
headline 308-request metric. Scoring exact key agreement as the only number
was rejected: ``the lightest possible part'' is
better answered by part density than by relative density, and a model that
says so should not be failed. Categories cover simple, compositional,
directional, contradictory, vocabulary, feasible, and infeasible requests.

\textbf{Is the output set reproducible?} The two stages answer
differently and should not be quoted as one number. Retrieval is
deterministic by construction: it is a boolean mask intersection followed
by stable sorts over a frozen table, so the same parse returns the same row
bit for bit, and the output set is finite ($1{,}397$ feasible geometries
$\times$ $19$ metals), enumerated rather than sampled, and closed --- every
query ends in one row or one attributed refusal. The language stage is the
only stochastic component. It runs at temperature $0$, which no provider
guarantees to be deterministic, so it is measured rather than assumed: two
full repeat runs of the $308$-request benchmark on the deployed Flash SKU
returned identical scores, and five runs on Flash-Lite spread
$1.95$ percentage points (sample standard deviation $0.7$). Run-to-run
variation therefore sits entirely in the parse, is small on the deployed
model, and cannot propagate into the search once the parse is fixed.

\begin{center}
\captionof{table}{Concept F1 on 308 template-generated requests. Flash-Lite is mean
$\pm$ sample standard deviation over five runs. Flash is two identical runs;
we do not write a five-run error bar. Keyword and TF-IDF are registry-derived
baselines, not learned parsers.}
\label{tab:parse}
\small
\begin{tabular}{@{}lcccc@{}}
\toprule
 & Flash-Lite & Flash & Keyword table & TF-IDF 1-NN \\
\midrule
Overall & $0.897\pm 0.007$ & $0.997$ & $0.606$ & --- \\
Literal & $1.000\pm 0.000$ & $1.000$ & --- & --- \\
Paraphrased & $0.793\pm 0.014$ & $0.994$ & $0.49$ & $0.57$ \\
\bottomrule
\end{tabular}
\end{center}

Gemini 3.5 Flash-Lite, five runs: overall $0.897\pm 0.007$, literal $1.000$,
paraphrased $0.793\pm 0.014$. Gemini 3.5 Flash, two runs, both $0.997$
overall and $0.994$ paraphrased. The keyword table generated from the same
registry scores $0.606$ overall. On held-out paraphrases, a TF-IDF 1-NN
trained on the literal half scores $0.57$ against the table's $0.49$ and
Flash-Lite's $0.80$.

The honest reading is that a mid-tier 2026 model saturates this vocabulary. A
paper whose contribution was ``an LLM beats a keyword table'' would be
answering a closed question. Flash-Lite remains a real deployment result: it
is sufficient on literal input and not on paraphrase. The parse stage is
therefore kept thin on purpose. Prompts are template-generated, not
human-authored. We do not claim robustness to unconstrained engineer prose;
Flash-Lite's $0.79$ paraphrase F1 is the available warning
(Sec.~\ref{sec:limits}). Only Gemini~3.5 Flash and Flash-Lite were run.

An ablation that varies the registry from two properties to the full set shows
that vocabulary size grows the LLM--table gap by about $0.11$, while
paraphrasing costs both about $0.21$ at full vocabulary. The table already
trails by $0.16$ at two properties. We do not claim that a keyword table is
optimal over a small property space.

\subsection{Min-repair as the comparison of record}
\label{sec:baselines}

Five search policies are scored on a frozen suite of 64 typed queries
(28 single-constraint thresholds, 20 pairs, 12 triples, and the four worked
briefs of Table~\ref{tab:briefs}). Feasibility labels are a direct
mask-intersection oracle computed before search, not each method's own row
list. A method's satisfy bit is then an objective comparison of its returned
row against those same constraints; it does not score a method by reusing the
system's mask code. MUS and MCS on empty queries are scored against an
independent enumerator (Sec.~\ref{sec:refusal}). The suite is systematic over registry keys, not independently authored
engineer prose.

\emph{Nearest neighbour} ignores hard constraints and ranks by the stated
objectives, or by rank-normalised slack when there is no objective.
\emph{Penalty search} maximises the same objective minus $\lambda=5$ times
normalised constraint violation and always returns a row. Both are the
behaviour a retrieval system defaults to when it must always return a row;
their constraint violations restate those definitions.

\emph{Min-repair} is the comparison of record because it is the constraint-aware
policy an engineer would actually build once refusal is on the table. If the
query is feasible, it returns the same rank-normalised row as
retrieve-or-refuse. If not, it takes a minimum-cardinality MCS---an
inclusion-minimal set of constraints whose removal restores a non-empty
feasible set---and returns the jointly attainable repair row on the remaining
mask. When several minimum-cardinality MCS exist, the first in the reasoner's
enumeration order is used; kept constraints are then satisfied by construction.
The reported repair values come from one row, so a multi-element MCS is a
jointly attainable point, not independent per-constraint extrema.
\emph{Lex-drop} peels constraints in a fixed property order
(\texttt{cost\_per\_kg}, \texttt{k\_aniso}, \texttt{mass\_density},
\texttt{rho}, \texttt{k\_11}, \texttt{E\_11}) until a row appears: same
always-answer habit as min-repair, coarser drop rule.

\begin{figure}[t]
\centering
\includegraphics[width=0.92\textwidth]{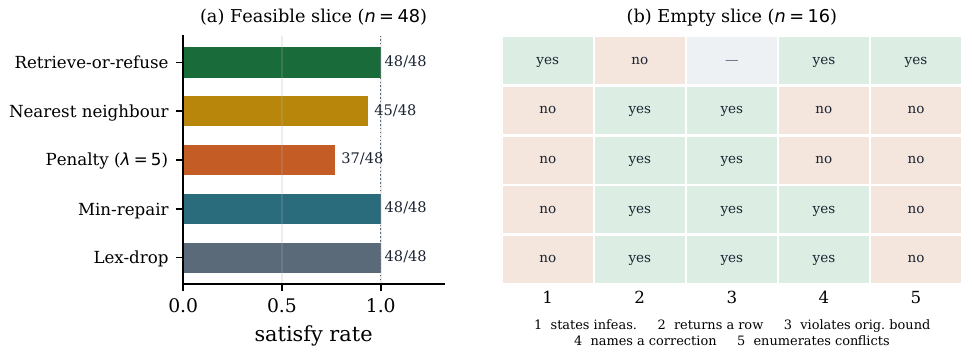}
\caption{Frozen typed suite ($n=64$; 48 feasible, 16 empty) against a
mask-intersection oracle computed before search. (a)~Fraction of feasible
queries whose returned row meets every original constraint. (b)~Empty queries:
comparable yes/no outcomes at the same printed precision, columns 1--5:
states infeasibility, returns a row, that row violates an original bound,
names a selected correction set, enumerates all minimal conflicts.
Retrieve-or-refuse
agrees with the oracle on all 64 decisions and matches an independent MUS
enumerator on the 16 empty queries. Min-repair satisfies every kept constraint
($64/64$). Median search $2.7$\,ms ($|\mathcal{C}|\le 3$).}
\label{fig:suite}
\end{figure}

Fig.~\ref{fig:suite} reports the suite.
Min-repair matches retrieve-or-refuse on every feasible query ($48/48$): both
select from the mask intersection, so that score is a correctness check, not
a ranking result. The measured difference is on the 16 empty queries:
retrieve-or-refuse refuses and returns MUS and slack; min-repair drops a
minimum-cardinality MCS and returns a jointly attainable row that satisfies
every kept constraint ($64/64$ over the whole suite) but none of the original
empty queries ($0/16$). Whether an engineer prefers a named refusal to one
selected correction set and its jointly attainable repair values is not
evidenced here; no user study was run. Nearest neighbour
satisfies $45/48$ feasible queries---the unconstrained optimum already lies
inside the feasible set on those---and $0/16$ infeasible ones. Penalty search
is worse on the feasible slice ($37/48$) because the penalty still trades a
hard limit for score. Lex-drop matches min-repair's feasible-set behaviour.
The deployed MUS matches an independent enumerator on all 16 empty queries.

\subsection{Printed-precision repair check}
\label{sec:repair-witness}

The suite comparison above scores each method on whether it returns a row
that meets the original constraints. It does not ask whether a reported
repair is executable after the numbers are printed. That question is
pre-registered on the 16 empty suite queries plus 200 additional empty
queries of two to five constraints (seed $20260905$; exact constraint-set
duplicates dropped; near-duplicate threshold none). Constraints that the
correction set does not drop keep their original values. Only repaired atoms
are printed at the tool's displayed precision (three significant figures,
round to nearest even---the former display) before search is re-run. This
experiment checks catalogue-level printed executability; it does not remesh
the selected cell. The catalogue row that attains a repair is the
\emph{witness}: the jointly attainable values against which a printed bound
is judged.

Five policies are named once here and used unchanged below.
\emph{First repair} takes the first minimum-cardinality MCS in reasoner
order. \emph{Best-objective repair} takes, among all minimum-cardinality MCS,
the jointly attainable row with the highest original-objective score
(rank-normalised, as in the rest of the paper; if the query has no
objective, the mean rank-normalised orientation toward the original
constraints). \emph{Full diagnosis} enumerates the MUS family and every
inclusion-minimal MCS. \emph{Smallest-first list with protection} ranks
globally smallest correction sets first, then rejects those that touch a
protected bound. \emph{Protection-first repair} enforces protection before
minimising the number of allowable relaxations, without MUS or MCS
enumeration. Lex-drop is not this baseline. None of the 216 queries has an
objective, so best-objective repair uses that fallback score throughout.
Executability below is scored for the two single-answer policies. Full
diagnosis is scored under the same action protocol in the revision task that
follows.

Acting on the printed repair restores a non-empty feasible set for
$107/216$ first-repair outputs and $94/216$ best-objective-repair
outputs. The usual failure is the displayed three-significant-figure bound
landing on the tight side of the jointly attainable value (the gold empty
query $\rho\le 0.15$, $E_{11}\ge 100$\,GPa prints $\rho\le 0.325$ under
nearest-even three-significant-figure rounding against a witness at
$\rho=0.32503$). Rounding repaired bounds outward at the same three
significant figures ($\le$ up, $\ge$ down), on the same frozen 216 queries,
restores feasibility for $216/216$ first-repair outputs and $216/216$
best-objective-repair outputs; the gold query then prints $\rho\le 0.326$.
That outward rule is what the deployed printer now uses. Nearest-even
three-significant-figure rounding is not a guaranteed feasible instruction.

The two single-answer methods disagree on the dropped set in $119/216$
queries and on the selected cell in $107/216$. Every one of those 119
dropped sets names a different property; of the 107 different cells, 53 are
a different metal and 80 a different family. Twelve queries share a cell and
still name a different constraint. Of the 119 set disagreements, 117 have
minimum cardinality one---a pairwise conflict always has two singleton
repairs---and two have minimum cardinality two. An inclusion-minimal MCS
that is not minimum-cardinality appears in $24/216$; that frequency is
additional information, not by itself a capability. Different dropped
properties and different cells show that repair-selection policy matters.
They do not by themselves show that full diagnosis outperforms a transparent
list of min-cardinality repairs.

The revision task was specified before its results. Density and unit cost,
when present, must keep their original bounds; stiffness, conductivity,
anisotropy, and part density may be loosened.
$211/216$ queries have at least one protected and one allowable constraint.
Smallest-first list with protection preserves those requirements on
$209/211$ queries.
The two misses are the pattern in Table~\ref{tab:mus}: $\rho$ is the unique
minimum MCS, and keeping $\rho$ requires dropping both $E_{11}$ and $k_{11}$.
Protection-first repair, a post-result control without MUS or MCS
enumeration, preserves $211/211$ and matches full diagnosis's repair loss on
every eligible query, including those two. Full diagnosis, allowed to use a
larger inclusion-minimal MCS, also preserves $211/211$. The $2/211$ gap is
the cost of retaining only globally smallest correction sets; it is not a
unique capability of enumerating the MUS family. First repair, which does
not use the protection mask, preserves $133/211$. On the 209 queries that
both smallest-first list with protection and full diagnosis preserve, mean
allowable constraints dropped is $1.01$ either way.

\begin{table}[htbp]
\centering
\caption{Printed-precision repair check ($n=216$ empty queries). The
nearest-even three-significant-figure rows are the pre-registered metric. The
outward rows are a follow-up display rule on the same frozen queries, not a
new draw. Retained original bounds are kept exactly; only repaired atoms are
printed. Executability is a non-empty catalogue search after substituting the
displayed repair values. Disagreement is over queries where both
single-answer methods return a repair. Policy names are those of
Sec.~\ref{sec:repair-witness}.}
\label{tab:repair-witness}
\begin{tabular}{@{}lcc@{}}
\toprule
& count & rate \\
\midrule
First repair executable at nearest-even 3\,s.f. & $107/216$ & $0.50$ \\
First repair executable, outward 3\,s.f.\ (same queries) & $216/216$ & $1.00$ \\
Best-objective repair executable at nearest-even 3\,s.f. & $94/216$ & $0.44$ \\
Best-objective repair executable, outward 3\,s.f. & $216/216$ & $1.00$ \\
Disagree on dropped set & $119/216$ & $0.55$ \\
\quad of which minimum cardinality one & $117/119$ & $0.98$ \\
Disagree on the returned cell & $107/216$ & $0.50$ \\
Among those, different metal & $53/107$ & $0.50$ \\
\bottomrule
\end{tabular}
\end{table}

\begin{table}[htbp]
\centering
\caption{Requirement-revision check on the same 216 queries. The task was
specified before its results. Density and unit cost are protected;
stiffness, conductivity, anisotropy, and part density may be loosened.
$211$ queries have at least one protected and one allowable constraint.
Smallest-first list with protection ranks globally smallest correction sets,
then rejects those that touch a protected bound. Protection-first repair
is a post-result control on the same frozen queries, not a new draw: it
enforces protection before minimising allowable relaxations, without MUS
or MCS enumeration. Preserves means the printed revised query is
executable and every protected original bound is still present.}
\label{tab:repair-revision}
\begin{tabular}{@{}lcc@{}}
\toprule
& count & rate \\
\midrule
Eligible queries & $211/216$ & $0.98$ \\
First repair (protection unused) preserves protected & $133/211$ & $0.63$ \\
Smallest-first list with protection & $209/211$ & $0.99$ \\
Protection-first repair (no MUS) & $211/211$ & $1.00$ \\
Full diagnosis (all inclusion-minimal MCS) & $211/211$ & $1.00$ \\
Smallest-first list fails; others succeed & $2/211$ & $0.01$ \\
\bottomrule
\end{tabular}
\end{table}

\subsection{A second transport property as a registry declaration}
\label{sec:dstar}

The same cells and the same scalar homogenisation operator, pointed at the
complete periodic pore, yield a dimensionless diffusivity $D^{\ast}/D_0$.
The pore definition is the complement of the stored-isovalue solid mask, not
the largest periodic labyrinth and not an inlet-accessible component. Three
declarations (\texttt{D\_11}, \texttt{D\_33}, \texttt{D\_aniso}) were added
to the registry. Search, MUS enumeration, and mask intersection were not
rewritten for $D^{\ast}$. (Displayed repair bounds were later rounded
outward; that printer change is Section~\ref{sec:repair-witness}, not a
reasoner change for $D^{\ast}$.) On a 16-query typed suite (8 feasible, 8 empty; two
pairwise conflicts and one triple), retrieve-or-refuse matches the
mask-intersection oracle on all 16 and min-repair satisfies every feasible
query ($8/8$). Thresholds are round values placed at stored quartiles:
$D_{11}\ge 0.40$,
$0.55$, $0.70$ sit at the lower quartile, above the median, and in the upper
quartile of stored $D_{11}$; $D_{\mathrm{aniso}}\le 0.70$ is the same round
value used for $k_{33}/k_{11}$; $D_{11}\ge 0.90$ lies above the stored
maximum $0.866$, and $D_{\mathrm{aniso}}\le 0.10$ below the stored minimum
$0.105$. High $D_{11}$ with high $\rho$, or with a tight $D_{33}$ or
$D_{\mathrm{aniso}}$ bound, is empty. This is a bounded schema extension, not
transfer to disjoint physics. No Archie's-law fit is claimed.

\subsection{Retrieve-or-refuse versus min-repair on worked briefs}
\label{sec:briefs}

The four briefs in Table~\ref{tab:briefs} are the worked cases. Their
numeric thresholds are stated here with their origin, because a threshold
chosen after seeing the data would make any brief succeed. None was: the
density caps come from the printable band established in
Sec.~\ref{sec:homog} --- below $\rho\approx 0.10$ the ligaments fall under
the powder-bed feature limit at a reasonable cell size, above
$\rho\approx 0.5$ a TPMS is a solid with holes --- and $0.40$ and $0.25$ sit
inside it with margin, in a density range used for heat-spreader
lattices. The price caps are round numbers just above the
bulk price of the obvious candidate metal in each case ($3$\,USD/kg above
aluminium at $2.5$, $5$ above it with headroom, $10$ above AlSi10Mg at $4$),
chosen so the cap binds against silver and copper rather than being
decorative. The anisotropy bound $k_{33}/k_{11}\le 0.70$ is the round value
below which no cubic cell can reach, so it forces the geometry to do the
work; it is the one threshold picked for what it excludes, and it excludes a
symmetry class rather than a data range.

One clarification, because the density caps invite it. The printable band
brackets $\rho$ but does not select $0.40$ over $0.30$, and neither number is
offered as a definition of ``light''. A cap is there so the request actually
limits the search: the returned row lands at the top of what the cap allows,
so removing the cap would change the answer. It is not a claim about where
lightness begins. Every value in $0.25$--$0.50$ produces the same demonstration ---
the whole span the catalogue can test --- because what the brief tests is whether
the search honours a stated limit and reports the trade, not whether the limit
is the correct one. The same reading applies to every other threshold
here. Where a number does carry physical content we say what fixes it --- the
printable floor near $\rho\approx 0.10$, the lattice ceiling near $0.50$, the
cubic bound at $k_{33}/k_{11}=0.70$ --- and where it does not, it is a round
number chosen so the constraint bites. That is testable rather than rhetorical.
Re-running the heat-spreader brief with the cap swept over
$\rho\le 0.25,\,0.30,\,\dots,\,0.50$, every other constraint held fixed, the cap
binds at each value: the returned row sits on it. Aluminium~6061 wins the
$5$\,USD/kg limit throughout, and the geometry changes at every step --- five
families across the six caps --- because geometry is what spends the density
budget. Only the position on the trade curve
moves, $k_{11}$ from $36.3$ to $77.5$\,W/mK; what the brief demonstrates does
not.

The
heat-spreader brief is the flagship: $k_{33}/k_{11}\le 0.70$, $\rho\le 0.40$,
cost at most $5$\,USD/kg, maximise $k_{11}$. No metal choice satisfies the
anisotropy bound; geometry has to do the work. Retrieve-or-refuse returns
aluminium 6061 diamond sheet $f{=}133$ at $\rho=0.400$, $k_{11}=59.8$\,W/mK
($3{,}168$ feasible). Nearest neighbour returns unconstrained silver; penalty
search returns copper and breaks the price cap.

The cheap-conductor brief is ``maximise $k_{11}$, material price $\le 3$\,USD/kg.''
Retrieve-or-refuse returns aluminium 6061, split-P network, $f{=}133$,
$\rho=0.509$, $k_{11}=78.2$\,W/mK, at $2.5$\,USD/kg ($4{,}191$ feasible rows).
Nearest neighbour returns silver at $900$\,USD/kg. Penalty search returns
copper at $9$\,USD/kg. Both always-answer rows violate the price constraint
that the registry compiled. Removing the price cap, the same search returns
silver at $k_{11}=201$\,W/mK: the material table is doing the work that a
geometry-only catalogue cannot do. Restricting the product to aluminium 6061
returns the same cell, because that metal is already the cheap conductor;
the contrast is not geometry-within-aluminium, it is silver versus aluminium
under the price cap.

The stiff-and-light brief (maximise specific stiffness,
$\rho\le 0.25$, cost $\le 10$\,USD/kg) returns AlSi10Mg Schwarz~P sheet
$f{=}133$ at $\rho=0.227$. Both constraint-ignoring methods return silicon
carbide and break the price cap.

On the empty light-and-stiff brief, min-repair returns tungsten diamond sheet
$f{=}133$ at $\rho=0.325$, $E_{11}=103$\,GPa: it keeps the stiffness bound and
names density as the repair. Nearest neighbour returns tungsten at
$\rho=0.257$, $E_{11}=81$\,GPa and violates both.

\begin{figure}[htbp]
\centering
\includegraphics[width=0.86\textwidth]{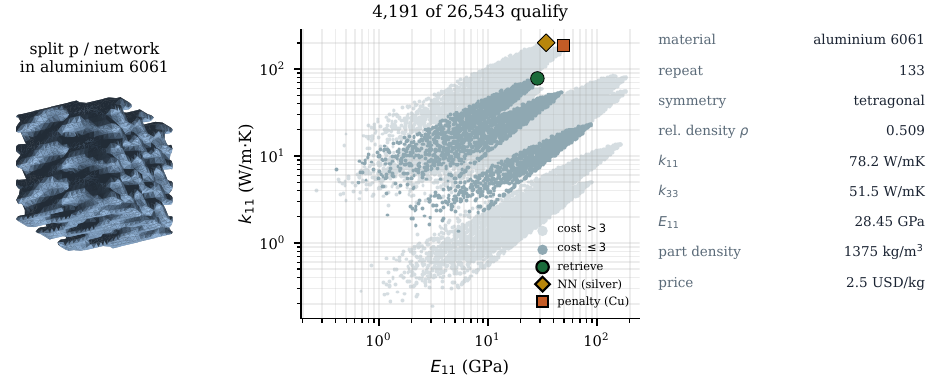}
\caption{Gold-query retrieval for maximum in-plane conductivity at bulk price
at most $3$\,USD/kg. Left: the returned cell. Centre: its location in the
$(E_{11},k_{11})$ cloud; the lighter points exceed the price cap. The diamond
is unconstrained nearest-neighbour (silver, $900$\,USD/kg); the square is
penalty search (copper, $9$\,USD/kg). Right: stored properties of the top row.}
\label{fig:result}
\end{figure}

Fig.~\ref{fig:result} is the cheap-conductor row in the design space, so the
figure tests search rather than parse.

\begin{table}[htbp]
\centering
\caption{Worked typed briefs. NN ignores constraints. Min-repair returns the
retrieve row when the query is feasible, and the jointly attainable MCS
repair when it is not.}
\label{tab:briefs}
\small
\begin{tabular}{@{}>{\raggedright\arraybackslash}p{0.28\textwidth}
  >{\raggedright\arraybackslash}p{0.33\textwidth}
  >{\raggedright\arraybackslash}p{0.33\textwidth}@{}}
\toprule
Brief & Retrieve-or-refuse & NN / min-repair \\
\midrule
Heat spreader\newline
$\max k_{11}$, $k_{33}/k_{11}{\le}0.7$, $\rho{\le}0.40$, cost $\le 5$
  & Al 6061 diamond sheet, $k_{11}{=}59.8$ ($n{=}3168$)
  & NN: silver (violates cost).\newline
    Min-repair: same as retrieve \\[4pt]
Cheap conductor\newline
$\max k_{11}$, cost $\le 3$
  & Al 6061 split-P, $k_{11}{=}78.2$, $2.5$\,USD/kg ($n{=}4191$)
  & NN: silver, $900$\,USD/kg (violates cost).\newline
    Min-repair: same as retrieve \\[4pt]
Stiff and light\newline
$\max E/\rho$, $\rho{\le}0.25$, cost $\le 10$
  & AlSi10Mg Schwarz~P sheet, $\rho{=}0.227$ ($n{=}3336$)
  & NN: SiC (violates cost).\newline
    Min-repair: same as retrieve \\[4pt]
Light and stiff\newline
$\rho{\le}0.15$, $E_{11}{\ge}100$
  & refuse; MUS $\{\rho,E_{11}\}$; slacks $0.326$ / $36.5$\,GPa
  & NN: W Schwarz~P, $\rho{=}0.257$, $E_{11}{=}81$ (violates both).\newline
    Min-repair: W diamond, $\rho{=}0.325$, $E_{11}{=}103$ (keeps $E_{11}$) \\
\bottomrule
\end{tabular}
\end{table}

\subsection{Both constraints can be feasible and their intersection empty}
\label{sec:refusal}

\begin{figure}[htbp]
\centering
\includegraphics[width=0.84\textwidth]{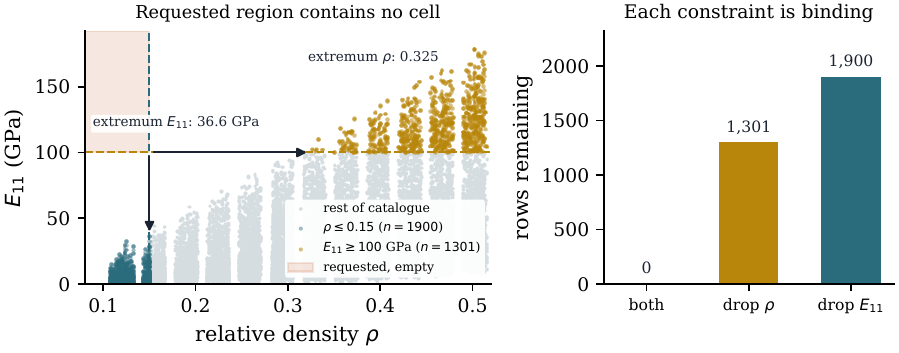}
\caption{Gold-query refusal of $\rho\le 0.15$ and $E_{11}\ge 100$\,GPa. Left:
the catalogue in $(\rho,E_{11})$. Teal points meet the density bound
($n=1{,}900$); gold points meet the stiffness bound ($n=1{,}301$). Their
intersection (shaded) is empty. Arrows are catalogue extrema, not the printed
slack: the lightest cell at $E_{11}\ge 100$\,GPa has $\rho=0.325$, and the
stiffest cell at $\rho\le 0.15$ has $E_{11}=36.6$\,GPa. The deployed printer
reports those repair bounds as $\rho\le 0.326$ and $E_{11}\ge 36.5$\,GPa
(Table~\ref{tab:briefs}). Right: rows remaining after dropping each singleton
MCS. Each constraint is feasible alone, so both are minimum-cardinality
repairs.}
\label{fig:refusal}
\end{figure}

Fig.~\ref{fig:refusal} is the gold query ``$\rho\le 0.15$ and
$E_{11}\ge 100$\,GPa.'' Nothing survives. The MUS is the pair
$\{\rho,E_{11}\}$; leave-one-out names both, which agrees because the unique
MUS is a pair. The slacks are the repairs: relative density would have to
reach $0.326$ instead of $0.15$, or stiffness $36.5$\,GPa instead of
$100$\,GPa. Table~\ref{tab:briefs} shows what always-answer does with the same
query: nearest neighbour returns tungsten at $\rho=0.257$, $E_{11}=81$\,GPa
(violates both); min-repair returns tungsten at $\rho=0.325$,
$E_{11}=103$\,GPa (keeps stiffness, repairs density); penalty search returns
tungsten at $\rho=0.317$, $E_{11}=99.5$\,GPa (meets stiffness to a percent,
breaks density by a factor of two, and presents it as an answer).

The same deployed diagnosis was scored on the 80-item constructed boundary
bench in Table~\ref{tab:refusebench} (no language model): 56 single-constraint
requests sitting $2$--$20\%$ either side of an achievable limit, and 24 jointly
infeasible pairs in which each constraint is feasible alone. The union of
minimum-cardinality MCS matches the constructed constraint set on every item,
including all 24 joint pairs (each pair is one MUS, so both singletons are
minimum MCS). That is a test of the diagnosis against how the requests were
built, not of whether a parser preserves feasibility, and not of higher-order
conflicts. An independent enumerator on five named conflict patterns and 72
random 1--6 constraint queries agreed with the deployed MUS and MCS on every
case, including a disjoint-MUS hitting-set identity on a four-element
synthetic family. Duplicate bounds on one property are distinct atoms: the
interval $\rho\le 0.20$ and $\rho\ge 0.30$ is one MUS of two $\rho$
constraints, two singleton MCS, and two different slacks. Serialising MUS by
property name would have collapsed both MCS to $\{\rho\}$ and printed the
wrong slack.

\begin{table}[htbp]
\centering
\caption{Union of minimum-cardinality MCS on the 80-item constructed boundary
bench. Items are typed from the boundary set; the language model
is out of the loop. This set is disjoint from the two worked gold queries
in Figs.~\ref{fig:result}--\ref{fig:refusal}.}
\label{tab:refusebench}
\small
\begin{tabular}{@{}lcccc@{}}
\toprule
Slice & $n$ & refuse & exact & set-F1 \\
\midrule
All & 80 & 56 & $80/80$ & $1.00$ \\
Single-constraint boundary & 56 & 32 & $56/56$ & $1.00$ \\
Jointly infeasible pairs & 24 & 24 & $24/24$ & $1.00$ \\
\bottomrule
\end{tabular}
\end{table}

Table~\ref{tab:mus} reports the cases that the 80-item set cannot see, now
from the deployed search rather than an offline script. A
three-constraint query with an extra non-binding limit
($\rho\le 0.15$, $E_{11}\ge 100$, $k_{11}\ge 40$) has MUS $\{\rho,E_{11}\}$;
leave-one-out correctly omits $k_{11}$. A query in which $\rho$ participates
in two pairwise conflicts
($\rho\le 0.18$, $E_{11}\ge 50$, $k_{11}\ge 60$) has MUS
$\{\{\rho,E_{11}\},\{\rho,k_{11}\}\}$; leave-one-out names only $\rho$, which
is the unique minimum MCS, and hides that there are two distinct pairwise
conflicts. The size-two MCS $\{E_{11},k_{11}\}$ is also inclusion-minimal: a
joint repair that keeps $\rho\le 0.18$ lands at $E_{11}=44.8$\,GPa and
$k_{11}=22.6$\,W/mK on one row, values that are not the independent extrema
of those two properties. A physically motivated three-way conflict---light,
stiff enough, and conducting enough
($\rho\le 0.20$, $E_{11}\ge 40$\,GPa, $k_{11}\ge 40$\,W/mK)---has every pair
non-empty ($n=67$, $287$, $2272$) and an empty triple. The MUS is the full
triple; leave-one-out names all three, which is correct as a hitting set and
silent on order. Always-answer search has no object corresponding to any of
these rows.

An always-refuse baseline at the 80-item set's infeasible base rate $0.70$
would score precision $0.70$, recall $1.00$, F1 $0.82$ on the refuse/answer
decision. Parse-then-search agreement with catalogue feasibility is high on
the template set; it is not reported as a recognition result, because both
sides call the same search (parse-feasibility agreement, not independent
recognition of impossibility).

\begin{table}[htbp]
\centering
\caption{Leave-one-out binding versus deployed inclusion-minimal unsatisfiable
subsets on three-constraint queries that are not in the 80-item constructed
set. Pair counts are feasible rows for each two-constraint projection. The
middle row is why MUS is not decoration: leave-one-out names the unique
minimum MCS and hides the two pairwise conflicts.}
\label{tab:mus}
\small
\begin{tabular}{@{}p{4.2cm}lll@{}}
\toprule
Query & LOO & MUS & Pair $n$ \\
\midrule
$\rho{\le}0.15$, $E_{11}{\ge}100$, $k_{11}{\ge}40$
  & $\rho,E_{11}$
  & $\{\rho,E_{11}\}$
  & extra $k_{11}$ not in MUS \\[3pt]
$\rho{\le}0.18$, $E_{11}{\ge}50$, $k_{11}{\ge}60$
  & $\rho$
  & $\{\rho,E_{11}\}$, $\{\rho,k_{11}\}$
  & two pairwise conflicts \\[3pt]
$\rho{\le}0.20$, $E_{11}{\ge}40$, $k_{11}{\ge}40$
  & $\rho,E_{11},k_{11}$
  & $\{\rho,E_{11},k_{11}\}$
  & $67$, $287$, $2272$ \\
\bottomrule
\end{tabular}
\end{table}

\subsection{On two gold briefs, refusal is stable under mesh-scale noise; top-1 rank is not}
\label{sec:stability}

A 60-cell mesh probe (Sec.~\ref{sec:mesh}) puts the median working-grid
residual at $1.6\%$ in $\kstar$ and $2.4\%$ in $\Estar$, so the $1\%$ draws
below are at the optimistic end of that spread rather than at its centre. The
next two paragraphs report two different tests, not one general stability
theorem.

On the two gold briefs, forty independent Gaussian perturbations of stored
$k_{11}$ and $E_{11}$ at relative scale $1\%$ leave the cheap-conductor
feasible set non-empty and the light-and-stiff set empty in every draw
(0/40 refuse/answer flips). The same holds at $5\%$. Top-1 identity on the
cheap-conductor brief is not stable: $27/40$ draws at $1\%$ change the
returned material--family--frequency triple, and $35/40$ at $5\%$.
The original winner remains inside the top-10 in $40/40$ draws at $1\%$
on that brief and on the heat-spreader brief. Rank-normalised search over a dense product is sensitive to mesh-scale noise;
the retrieve-or-refuse decision on those two briefs is not. A returned top
row should be read as a member of that stable top-10, not as a unique
optimum.

On the 80-item constructed boundary set the test is stricter: each item
perturbs the effective properties named in its gold query at $1\%$, twenty
draws per item ($1{,}600$ searches). Refuse/answer flipped in $2/1600$ draws,
on two of 56 single-constraint boundary items and on none of the 24 joint
pairs. Those two items sit close to an achievable limit by construction; the
gold briefs do not. The refusal claim is therefore that far-from-boundary
queries and most constructed boundary queries keep their decision under
mesh-scale noise, not that every query does.

\section{Discussion}
\label{sec:discuss}

The two properties claimed in Sec.~\ref{sec:intro} are what survive contact
with the literature and the suite. A keyword table cannot represent ``cheap enough to print,
conduct sideways, insulate upward'' as a coupled query over metals and cells.
A geometry-only library cannot move $k/E$ by $94\times$.
A system that always returns five neighbours cannot tell an engineer that
$\rho\le 0.15$ and $E_{11}\ge 100$\,GPa do not coexist in the catalogue, or by
how much.

Against MetaGen~\cite{makatura2025metagen} and TrussGPT~\cite{lin2026trussgpt}
the distinction is generate versus retrieve, and always-answer versus refuse.
Their contribution is a new geometry. Ours is an auditable row, or an
explanation. Against Wan~\emph{et al.}~\cite{wan2025aei} the distinction is
passage retrieval for Q\&A versus typed constraint search that can come back
empty. Against Chen and Bao~\cite{chen2026aei} and Liang~\emph{et
al.}~\cite{liang2025aei} the distinction is effort-measured generation and
optimisation versus a catalogue reasoner that names a minimal conflict.
Against Vyas~\emph{et al.}~\cite{vyas2026aei} the distinction is that they
measure designers and we do not; the human revision loop in
Fig.~\ref{fig:pipeline} is an interface claim, not a user study. Against OptiChat~\cite{chen2025optichat} and
MOID~\cite{li2025moid}, infeasibility diagnosis in natural language exists;
MOID already returns multiple revision suggestions from trade-off solutions.
Neither attaches that diagnosis to a lattice catalogue whose rows can be
re-solved.

A $1{,}397$-cell catalogue is a design choice. Every cell is a solve; none is
a prediction. That choice loses a scale comparison to~\cite{wang2026natcomm}
by two orders of magnitude, and it is the reason a returned number can be
rebuilt and re-solved. The paper is not a dataset contribution.

This paper answers the catalogue query; it does not qualify a printed part.
Nominal wrought constants are ranking values. In an early-lattice meeting the
change is operational: if the tool returns a row, that row can be rebuilt from
stored parameters; if
it returns a MUS and a slack, the next question is which requirement to
loosen, not which neighbour to pretend is an answer.

\subsection{Operational consequences}
\label{sec:practice}

Three operational rules follow from the evaluation, short of a user study.
First, on the two gold briefs, trust a returned row as a member of a stable
top-10 under $1\%$ mesh-scale noise, not as a unique optimum: the original
winner stayed inside the top-10 in every $1\%$ draw, while top-1 identity
flipped on most cheap-conductor draws. Rebuild the mask from the stored
parameters and re-homogenise before a drawing number enters a part. Second,
when the tool refuses, say which requirements are protected before choosing a
repair: with density and cost protected, protection-first repair kept them on
$211/211$ queries, as did full diagnosis, while first repair without that
information kept them on $133/211$. The MUS family is what to read before
the protection is known: it names every conflict, and two equally minimal
repairs differed in $119/216$ queries and in $107/216$ returned cells
(Table~\ref{tab:repair-witness}). Smallest-first list with protection keeps
those bounds on $209/211$ eligible queries
(Table~\ref{tab:repair-revision}). The $2/211$ tail is an ablation of that
list policy, not a unique MUS capability. The printed slack at three
significant figures, rounded outward, is a feasible bound on that frozen set.
Do not treat a neighbour that still violates the brief as a repair.
Third, read
\texttt{conn\_frac} on any returned row. An insulation query that minimises
$k_{11}$ can return a disconnected cell: \texttt{conn\_frac} is stored but is
not a queryable registry key, so the reasoner cannot take a connectivity floor.
Thirty-one searchable rows are more than $99\%$
disconnected solid, and four of the five lowest-$k_{11}$ product rows sit
below $3\%$ connected. Those rows are disclosed; they are not a recommended
answer.

\section{Limitations}
\label{sec:limits}

Every quoted effective property is computed. Material constants are nominal
wrought values; AM parts, temper, and build direction move them by tens of
percent. Elastic factorisation is computed at $\nu=0.3$ throughout. On sixteen
family--mode cells this shifts which cell is selected not at all over
$\nu\in[0.29,0.34]$ and by $0.40\%$ of $E_{11}$ at $\nu=0.45$, but it moves
the reported values themselves by up to $11.8\%$ in $E_{11}$ and $33.5\%$ in
$E_{33}$ across the full physical range. Absolute stiffness values are
therefore quoted with confidence only for materials near $\nu=0.3$; the
probe is sixteen cells at one density, not a catalogue-wide sweep of every
frequency. Multi-objective ranking is a weighted sum; one objective can
dominate. The Pareto count reports how many non-dominated alternatives remain;
it does not detect weight collapse. Top-1
rank on the cheap-conductor brief is unstable under $1\%$ property noise
($27/40$ family flips); refusal on that brief and on the empty gold query did
not flip at $1\%$ or $5\%$. The original cheap-conductor and heat-spreader
winners stay inside the top-10 in $40/40$ draws at $1\%$. On the 80-item constructed boundary set,
refuse/answer flipped in $2/1600$ draws. Mesh residual has a median of $1.6\%$ in
$\kstar$ and $2.4\%$ in $\Estar$ over a balanced 60-cell probe against $n=64$,
with a tail reaching $21.1\%$; all thirteen cells above $3\%$ sit at $\rho=0.25$,
across every frequency vector. Of the $1{,}397$ searchable geometries,
$31$ have largest-connected-solid fraction below $0.01$; \texttt{conn\_frac}
is stored but not queryable, so an insulation query that minimises $k_{11}$
can return those rows. The cubic identity $(k_{33}/k_{11},E_{33}/E_{11})=(1,1)$ holds up
to discretisation (worst $k_{33}/k_{11}=0.897$ at $n=32$). The parse benchmark is
template-generated; we do not claim robustness to unconstrained engineer
prose, and Flash-Lite's $0.79$ paraphrase F1 is the available warning. No
practising-engineer language study and no controlled human-use study were
run~\cite{vyas2026aei}. Independently authored engineer prose was not collected.
Printed repair values use three significant figures rounded outward in signed
value so a displayed bound cannot exclude the witness, including on strict
operators and on negative numbers. Constraints that are not repaired are not
re-rounded. That rule restores feasibility on $216/216$ frozen empty queries
for first repair. Nearest-even three-significant-figure rounding, the
pre-registered display, restores $107/216$. A minimal follow-up would measure, per refusal, whether
an engineer accepts the MUS, loosens a named constraint by the reported slack,
or abandons the brief. The registry now includes a second transport property
on the complete periodic pore ($D_{11}$, $D_{33}$, $D_{33}/D_{11}$); the
reasoner did not change. That is a bounded schema extension, not a transfer
to disjoint physics. Cross-implementation agreement verifies two codes on one
discretisation, not the physical cell. Flow is out of scope: a Kozeny--Carman
estimate exists in the code and is not claimed. A single vendor's language
models were used; the parse stage is specified so that an open-weights
replacement is a prompt change. MUS enumeration is capped at eight
constraints; larger empty queries are refused with that cap named. No
containment comparison against a prior system was run: of the closest
peers, the one whose output overlaps ours releases no code, and the one
that releases code reports a different output quantity, so the comparison
is left to future work rather than approximated.

\section{Conclusion}
\label{sec:concl}

Two properties are what this paper claims, and the suite is the evidence.
When the feasible set is empty the search returns why---a named MUS and the
slack of a repair---not the nearest row. A quantity the registry does not
declare cannot reach the evaluator as though it did. The language model does
not retrieve. The reasoner does not invent a neighbour. Against
constraint-aware min-repair, on the frozen typed suite, the feasible rows
coincide ($48/48$); the difference is that retrieve-or-refuse refuses the 16
empty queries with MUS and slack, while min-repair returns one selected
correction set and its jointly attainable repair values.
Constraint-ignoring nearest-neighbour and penalty search, kept
as reference, return rows that violate constraints the registry compiled.
On 216 frozen empty queries, outward three-significant-figure repairs are
executable for first repair and for best-objective repair ($216/216$). With
density and cost protected, smallest-first list with protection preserves
those bounds on $209/211$ queries; protection-first repair matches full
diagnosis at $211/211$. The lattice
catalogue is the instantiation that makes those two properties testable.

\section*{CRediT authorship contribution statement}
Shaoliang Yang: methodology, software, formal analysis, writing -- original
draft.\\
Henry Chu: validation, writing -- review and editing.\\
Zu Yashengjiang: validation.\\
Jun Wang: conceptualization, supervision, writing -- review and editing.

\section*{Data availability}
The catalogue, property registry, two worked gold queries, constructed
boundary bench, typed briefs, frozen suite, MUS/MCS oracle tests, Poisson
sweep, rank and refusal stability draws, sampled re-solve reports, the
frozen empty-query repair and revision checks, and the second-library
\texttt{scikit-fem}/TetGen logs used
in this manuscript will be released at
\url{https://github.com/nbbllxx0/A-PROPERTY-REGISTRY-CONTRACT-FOR-RETRIEVE-OR-REFUSE-THERMAL-MECHANICAL-LATTICE-SEARCH}
when the arXiv preprint is online.
Language-model prompts are generated from the registry. Appendices A--D
reproduce the full registry, the generated prompt, the 64-query suite, and one
tool transcript.

\section*{Declaration of generative AI and AI-assisted technologies in the writing process}
During the preparation of this work the authors used Cursor for language
editing of the manuscript. After using this tool, the authors reviewed and
edited the content as needed and take full responsibility for the content of
the publication. Language-model use in query parsing is a method of the
study, reported in the Evaluation, and is not writing assistance.

\appendix

\section{Full property-registry contract}

\label{app:registry}

\begin{table}[htbp]
\centering
\caption{Full property-registry contract. Table~\ref{tab:registry} in the main text is an excerpt.}
\label{tab:registry-full}
\scriptsize
\setlength{\tabcolsep}{2.8pt}
\begin{tabular}{@{}lllll@{}}
\toprule
Key & Kind & Unit & Provenance & Operators \\
\midrule
\texttt{rho} & geometry & --- & solved (isovalue bisection) & $\le,\ge$ \\
\texttt{porosity} & geometry & --- & solved ($1-\rho$) & $\le,\ge$ \\
\texttt{k\_aniso} & geometry & --- & solved ($k_{33}/k_{11}$, smaller is more directional) & $\le,\ge$ \\
\texttt{E\_aniso} & geometry & --- & solved ($E_{33}/E_{11}$) & $\le,\ge$ \\
\texttt{D\_11} & geometry & --- & solved ($D^{\ast}/D_0$, complete periodic pore) & $\le,\ge$ \\
\texttt{D\_33} & geometry & --- & solved ($D^{\ast}/D_0$, complete periodic pore) & $\le,\ge$ \\
\texttt{D\_aniso} & geometry & --- & solved ($D_{33}/D_{11}$) & $\le,\ge$ \\
\texttt{symmetry} & geometry & --- & solved (point group of $f$) & $=$ \\
\texttt{cost\_per\_kg} & material & USD/kg & handbook & $\le,\ge$ \\
\texttt{cte} & material & $10^{-6}$/K & handbook & $\le,\ge$ \\
\texttt{tmax} & material & $^\circ$C & handbook & $\le,\ge$ \\
\texttt{printable} & material & --- & handbook & $=$ \\
\texttt{k\_11} & effective & W/(m\,K) & solved $\times$ handbook, exact & $\le,\ge$ \\
\texttt{k\_22} & effective & W/(m\,K) & solved $\times$ handbook, exact & $\le,\ge$ \\
\texttt{k\_33} & effective & W/(m\,K) & solved $\times$ handbook, exact & $\le,\ge$ \\
\texttt{k\_mean} & effective & W/(m\,K) & solved $\times$ handbook, exact & $\le,\ge$ \\
\texttt{E\_11} & effective & GPa & solved $\times$ handbook, $\nu{=}0.3$ & $\le,\ge$ \\
\texttt{E\_22} & effective & GPa & solved $\times$ handbook, $\nu{=}0.3$ & $\le,\ge$ \\
\texttt{E\_33} & effective & GPa & solved $\times$ handbook, $\nu{=}0.3$ & $\le,\ge$ \\
\texttt{E\_mean} & effective & GPa & solved $\times$ handbook, $\nu{=}0.3$ & $\le,\ge$ \\
\texttt{mass\_density} & effective & kg/m$^3$ & handbook $\times$ $\rho$ & $\le,\ge$ \\
\texttt{specific\_stiffness} & effective & GPa/(kg/m$^3$) & solved $\times$ handbook & $\le,\ge$ \\
\texttt{specific\_conductivity} & effective & W/(m\,K)/(kg/m$^3$) & solved $\times$ handbook & $\le,\ge$ \\
\texttt{cost\_per\_m3} & effective & USD/m$^3$ & handbook $\times$ $\rho$ & $\le,\ge$ \\
\texttt{permeability} & effective & m$^2$ & estimated, not claimed & --- \\
\texttt{min\_feature} & effective & mm & estimated ($C\rho/S_v$), not claimed & $\le,\ge$ \\
\bottomrule
\end{tabular}
\end{table}

\section{Generated language-model prompt}

\label{app:prompt}

The generated system prompt in use at evaluation is reproduced below. It is produced by \texttt{prompt\_block()} in the registry; the text is not hand-edited.

\begin{lstlisting}[style=tool]
You translate an engineer's request for a porous metal part into a
structured query. You do not answer the request or suggest a material yourself
-- a physics search does that afterwards.

Axis convention: axes 1 and 2 are in-plane (sideways); axis 3 is through-thickness (up). 'along the length' or 'in-plane' means axis 1; 'across' or 'through' or 'upward' means axis 3.

Available properties:

[effective properties]
  k_11 (W/(m K)): conductivity along axis 1  -- heat conduction sideways, in-plane.
  k_22 (W/(m K)): conductivity along axis 2  -- the other in-plane direction.
  k_33 (W/(m K)): conductivity along axis 3  -- through-thickness conduction. Minimise this to insulate in one direction while conducting in the others.
  k_mean (W/(m K)): mean conductivity  -- average over the three axes.
  E_11 (GPa): stiffness along axis 1  -- Young's modulus in-plane.
  E_22 (GPa): stiffness along axis 2
  E_33 (GPa): stiffness along axis 3  -- through-thickness stiffness.
  E_mean (GPa): mean stiffness  -- average over the three axes. Registered because k_mean is: without it, a request like 'stiffness above 1 GPa' with no axis named parses to E_mean, fails validation, and the constraint is dropped -- which returns designs that violate it instead of refusing.
  mass_density (kg/m3): part density  -- actual mass per unit volume of the porous part, not the metal.
  specific_stiffness (GPa/(kg/m3)): stiffness per unit mass  -- the figure of merit when a part must be stiff and light.
  specific_conductivity (W/(m K)/(kg/m3)): conduction per unit mass  -- the figure of merit when a part must conduct and be light.
  cost_per_m3 (USD/m3): material cost per unit volume  -- price of the metal actually used; porosity makes a part cheaper.
  permeability (m2): permeability  -- how easily fluid flows through the pores, for a given cell size. ESTIMATE ONLY -- from porosity and surface area, not a flow solve.
  min_feature (mm): thinnest wall or ligament  -- how thin the metal gets, at the chosen cell size (default 10 mm). Ask for this to be at least 0.2 mm for a printable TPMS wall, or at least 0.4 mm for a comfortable powder-bed margin. Scaling the cell up scales this with it. ESTIMATE ONLY -- a mean thickness from density and surface area, not a measured minimum.

[geometry properties]
  rho: relative density  -- fraction of the cube that is solid, 0.12 to 0.50. Lower is lighter.
  porosity: porosity  -- fraction that is empty space; 1 minus relative density.
  k_aniso: k33/k11, through-thickness over in-plane conductivity  -- Ratio, not a magnitude of anisotropy. 1.0 means heat travels equally in all directions. Below 1 means it moves better sideways (axes 1,2) than through-thickness (axis 3), so SMALLER means more directional: 0.2 conducts five times better sideways than upward. A request to spread heat sideways and insulate upward asks for this to be small.
  E_aniso: E33/E11, through-thickness over in-plane stiffness  -- Ratio, not a magnitude of anisotropy. 1.0 means equally stiff in all directions; smaller means stiffer in-plane than through-thickness.
  D_11: pore diffusivity along axis 1  -- effective solute diffusivity D*/D0 of the complete periodic pore, in-plane. Dimensionless. High means the void conducts tracer well sideways.
  D_33: pore diffusivity along axis 3  -- through-thickness pore diffusivity D*/D0 of the complete periodic pore. Minimise to restrict tracer along axis 3.
  D_aniso: D33/D11, through-thickness over in-plane pore diffusivity  -- Ratio. 1.0 means the pore transports equally in all directions. Below 1 means tracer moves better sideways than through-thickness, so SMALLER means more directional.
  symmetry: symmetry class  -- cubic, tetragonal or orthorhombic. Cubic cells cannot steer heat at all -- their conductivity is identical in every direction.

[material properties]
  cost_per_kg (USD/kg): material price  -- approximate bulk price of the base metal.
  cte (1e-6/K): thermal expansion  -- how much it grows when heated. Low matters when parts must stay dimensionally stable.
  tmax (C): max service temperature
  printable: additively manufacturable  -- whether this metal is routinely 3D printed.

Rules:
- Use only property keys from the list above. Never invent one.
- A wish ("as light as possible", "cheap") is an objective. A requirement
  ("at least 5 GPa", "under 200 C") is a constraint with a number.
- If the request implies a direction, pick the axis-specific property rather
  than the mean.
- Convert units to those listed. Conductivity in W/(m K), stiffness in GPa.
- If the request mentions something this vocabulary cannot express -- fatigue,
  corrosion, cost of manufacture, anything absent from the list -- put it in
  `unmet` rather than forcing it into a property that does not mean the same
  thing.
- Requests can be contradictory or impossible. Translate them faithfully
  anyway; deciding feasibility is the search's job, not yours.
- Give every objective a positive `weight`. Use 1.0 when the request treats its
  wishes as equally important, and a larger number for whatever it emphasises
  ("mainly cheap, and if possible also stiff" -> cost 2.0, stiffness 1.0).
  Ordering carries no meaning; only the weights are read.
\end{lstlisting}

\section{Frozen 64-query suite}

\label{app:suite}

\begin{center}
\scriptsize
\setlength{\tabcolsep}{3pt}
\begin{longtable}{@{}p{3.15cm}cccccc@{}}
\caption{Frozen suite of 64 typed queries. Feas.\ is the mask-intersection label. R = retrieve-or-refuse (answered and satisfies); NN / Pen / MR / LD are whether that policy's returned row satisfies the original constraints. Y/N. Retrieve-or-refuse leaves empty queries unanswered.}\\
\label{tab:suite64}\\
\toprule
Query id & Feas.\ & R & NN & Pen & MR & LD \\
\midrule
\endfirsthead
\toprule
Query id & Feas.\ & R & NN & Pen & MR & LD \\
\midrule
\endhead
\bottomrule
\endfoot
\texttt{s\_rho\_<=\_0.12} & Y & Y & Y & Y & Y & Y \\
\texttt{s\_rho\_<=\_0.15} & Y & Y & Y & Y & Y & Y \\
\texttt{s\_rho\_<=\_0.18} & Y & Y & Y & Y & Y & Y \\
\texttt{s\_rho\_<=\_0.2} & Y & Y & Y & Y & Y & Y \\
\texttt{s\_rho\_<=\_0.25} & Y & Y & Y & Y & Y & Y \\
\texttt{s\_rho\_<=\_0.3} & Y & Y & Y & Y & Y & Y \\
\texttt{s\_rho\_<=\_0.4} & Y & Y & Y & Y & Y & Y \\
\texttt{s\_E\_11\_>=\_20} & Y & Y & Y & Y & Y & Y \\
\texttt{s\_E\_11\_>=\_40} & Y & Y & Y & Y & Y & Y \\
\texttt{s\_E\_11\_>=\_50} & Y & Y & Y & Y & Y & Y \\
\texttt{s\_E\_11\_>=\_80} & Y & Y & Y & Y & Y & Y \\
\texttt{s\_E\_11\_>=\_100} & Y & Y & Y & Y & Y & Y \\
\texttt{s\_E\_11\_>=\_150} & Y & Y & Y & Y & Y & Y \\
\texttt{s\_k\_11\_>=\_10} & Y & Y & Y & Y & Y & Y \\
\texttt{s\_k\_11\_>=\_40} & Y & Y & Y & Y & Y & Y \\
\texttt{s\_k\_11\_>=\_60} & Y & Y & Y & Y & Y & Y \\
\texttt{s\_k\_11\_>=\_80} & Y & Y & Y & Y & Y & Y \\
\texttt{s\_k\_11\_>=\_120} & Y & Y & Y & Y & Y & Y \\
\texttt{s\_cost\_per\_kg\_<=\_3} & Y & Y & Y & N & Y & Y \\
\texttt{s\_cost\_per\_kg\_<=\_5} & Y & Y & Y & N & Y & Y \\
\texttt{s\_cost\_per\_kg\_<=\_10} & Y & Y & Y & N & Y & Y \\
\texttt{s\_cost\_per\_kg\_<=\_40} & Y & Y & Y & Y & Y & Y \\
\texttt{s\_k\_aniso\_<=\_0.55} & Y & Y & Y & Y & Y & Y \\
\texttt{s\_k\_aniso\_<=\_0.7} & Y & Y & Y & Y & Y & Y \\
\texttt{s\_k\_aniso\_<=\_0.9} & Y & Y & Y & Y & Y & Y \\
\texttt{s\_mass\_density\_<=\_800} & Y & Y & Y & Y & Y & Y \\
\texttt{s\_mass\_density\_<=\_1200} & Y & Y & Y & Y & Y & Y \\
\texttt{s\_mass\_density\_<=\_2000} & Y & Y & Y & Y & Y & Y \\
\texttt{p\_00} & N & N & N & N & N & N \\
\texttt{p\_01} & N & N & N & N & N & N \\
\texttt{p\_02} & Y & Y & Y & Y & Y & Y \\
\texttt{p\_03} & N & N & N & N & N & N \\
\texttt{p\_04} & N & N & N & N & N & N \\
\texttt{p\_05} & Y & Y & Y & Y & Y & Y \\
\texttt{p\_06} & N & N & N & N & N & N \\
\texttt{p\_07} & Y & Y & Y & Y & Y & Y \\
\texttt{p\_08} & Y & Y & Y & Y & Y & Y \\
\texttt{p\_09} & Y & Y & Y & Y & Y & Y \\
\texttt{p\_10} & N & N & N & N & N & N \\
\texttt{p\_11} & Y & Y & Y & N & Y & Y \\
\texttt{p\_12} & Y & Y & Y & N & Y & Y \\
\texttt{p\_13} & Y & Y & Y & Y & Y & Y \\
\texttt{p\_14} & Y & Y & Y & Y & Y & Y \\
\texttt{p\_15} & Y & Y & Y & Y & Y & Y \\
\texttt{p\_16} & Y & Y & Y & Y & Y & Y \\
\texttt{p\_17} & Y & Y & Y & N & Y & Y \\
\texttt{p\_18} & Y & Y & Y & Y & Y & Y \\
\texttt{p\_19} & Y & Y & Y & Y & Y & Y \\
\texttt{t\_00} & N & N & N & N & N & N \\
\texttt{t\_01} & N & N & N & N & N & N \\
\texttt{t\_02} & N & N & N & N & N & N \\
\texttt{t\_03} & N & N & N & N & N & N \\
\texttt{t\_04} & N & N & N & N & N & N \\
\texttt{t\_05} & Y & Y & Y & N & Y & Y \\
\texttt{t\_06} & Y & Y & Y & Y & Y & Y \\
\texttt{t\_07} & N & N & N & N & N & N \\
\texttt{t\_08} & N & N & N & N & N & N \\
\texttt{t\_09} & Y & Y & Y & N & Y & Y \\
\texttt{t\_10} & N & N & N & N & N & N \\
\texttt{t\_11} & N & N & N & N & N & N \\
\texttt{r\_cheap\_k} & Y & Y & N & N & Y & Y \\
\texttt{r\_spreader} & Y & Y & N & N & Y & Y \\
\texttt{r\_stiff\_light} & Y & Y & N & N & Y & Y \\
\texttt{r\_light\_stiff} & N & N & N & N & N & N \\
\end{longtable}
\end{center}

\section{Worked tool transcript}

\label{app:transcript}

The empty light-and-stiff brief of Table~\ref{tab:briefs}, compiled as a typed query (no language-model parse). The print format is that of the deployed search loop.

\begin{lstlisting}[style=tool]
REQUEST   "rho <= 0.15 and E_11 >= 100 GPa"
------------------------------------------------------------------------------
UNDERSTOOD AS  rho <= 0.15; E_11 >= 100
               parse (typed query; no language model) | search | 26,543 combinations considered
------------------------------------------------------------------------------
NO CATALOGUE ROW SATISFIES THIS
  MUS: {rho<=0.15, E_11>=100}
  min MCS: {rho<=0.15}; {E_11>=100}
  closest achievable: rho<=0.15 would have to reach 0.326 instead of 0.15; E_11>=100 would have to reach 36.5 GPa instead of 100 GPa
\end{lstlisting}

\bibliographystyle{unsrtnat}
\bibliography{refs}
\end{document}